\documentclass[iop]{emulateapj}

\newcommand{\Msolar}{M$_{\odot}\,$}
\newcommand{\kms}{km s$^{-1}$}
\newcommand{\eoi}{$e/i$}

\newcommand{\SM}{349}     
\newcommand{\SN}{291}     
\newcommand{\BM}{78}      
\newcommand{\BN}{15}      
\newcommand{\BLM}{46}     
\newcommand{\BLN}{35}     
\newcommand{\BU}{28}      
\newcommand{\U}{204}      

\newcommand{\tot}{8252}   
\newcommand{\tRV}{1046}   
\newcommand{\tm}{473}     
\newcommand{\tvm}{124}    
\newcommand{\ttRV}{848}   
\newcommand{\orb}{93}     

\newcommand{\magn}{12$\leq$V$\leq$16.5}         
\newcommand{\mrv}{-42.36 $\pm$~0.04 \kms} 
\newcommand{\pr}{0.4 \kms}               
\newcommand{\vderr}{0.64 $\pm$ 0.04 \kms}               

\begin{document}

\title{WIYN Open Cluster Study XXXII: Stellar Radial Velocities in the Old Open Cluster NGC 188}
\shorttitle{WOCS: Stellar Radial Velocities in NGC 188}

\author{Aaron M. Geller\footnote{Visiting Astronomer, Kitt Peak National Observatory, National Optical Astronomy Observatory, which is operated by the Association of Universities for Research in Astronomy (AURA) under cooperative agreement with the National Science Foundation.}~~and Robert D. Mathieu,}
\affil{Department of Astronomy, University of Wisconsin - Madison, WI 53706 USA}

\author{Hugh C. Harris,}
\affil{United States Naval Observatory, Flagstaff, AZ, 86001, USA}

\author{Robert D. McClure}
\affil{Dominion Astrophysical Observatory - Herzberg Institute of Astrophysics - National Research Council \\ 5071 W. Saanich Road, Victoria, BC V9E 2E7 Canada}

\shortauthors{Geller et al.}

\begin{abstract}

We present the results of our ongoing radial-velocity survey of the old (7 Gyr) open cluster NGC 188.  
Our WIYN 3.5m data set spans a time baseline of eleven years, a magnitude range 
of \magn~(1.18 - 0.94 \Msolar), and a 1\degr~diameter region on the sky.  With the addition of a Dominion Astrophysical Observatory (DAO) 
data set we extend our bright limit to $V$ = 10.8 and, for some stars, extend our time
baseline to thirty-five years.   
Our magnitude limits include solar-mass main-sequence stars, subgiants, giants, and blue stragglers, and 
our spatial coverage extends radially to 17 pc ($\sim$13 core radii). 
For the WIYN data we present a detailed description of our data reduction process and a thorough analysis 
of our measurement precision of \pr~for narrow-lined stars.
We have measured radial velocities for \tRV~stars in the direction of NGC 188, and have calculated radial-velocity 
membership probabilities for stars with $\geq$3 measurements, 
finding \tm~to be likely cluster members.   We detect \tvm~velocity-variable cluster members,
all of which are likely to be dynamically hard binary stars.  
Using our single member stars, we find an average cluster radial velocity of \mrv.
We use our precise radial-velocity and proper-motion membership data to greatly reduce field star contamination in 
our cleaned color-magnitude diagram, from which we identify six stars of note that lie far from a standard single-star 
isochrone.  
We present a detailed study of the spatial distribution of cluster member 
populations, and find the binaries to be centrally concentrated, providing evidence for the presence of mass segregation in 
NGC 188.  We observe the blue stragglers to populate a bimodal spatial distribution that is not centrally concentrated,
suggesting that we may be observing two populations of blue stragglers in NGC 188, including a centrally concentrated 
distribution as well as a halo population. Finally, we find NGC 188 to have a global
radial-velocity dispersion of \vderr, which may be inflated by up to 0.23 \kms~from unresolved binaries.
When corrected for unresolved binaries, the NGC 188 radial-velocity dispersion has a nearly 
isothermal radial distribution.  We use this mean corrected velocity dispersion to derive a virial mass of 2300 $\pm$ 460 \Msolar.

\end{abstract}

\keywords{(galaxy:) open clusters and associations: individual (NGC 188) - (stars:) binaries: spectroscopic - (stars:) blue stragglers}

\section{Introduction}

As one of the richest old open clusters in our Galaxy, NGC 188 has been extensively studied.
The breadth of available data, old age, and the abundance of short-period photometrically variable stars and 
X-ray sources, make NGC 188 a very attractive 
cluster for both observational and theoretical astronomers.  Moreover, the  
multitude of photometric surveys \citep{san62,egg69,mcc74,mcc77,twa78,von98,sar99,pla03,ste04}, 
spectroscopic studies \citep{har85b,hob90}, short-period variable star 
studies \citep{efr64,van84,kal87,mac91,zha02,bel98,kaf03,gon05}, including both optical and X-ray photometric monitoring,
proper-motion (PM) analyses \citep{din96,pla03}, and the binary population study by \citet{mat04} 
provide an unprecedented basis from which to study an 
open cluster.  However, until now, the study of this cluster has lacked a critical component.
This paper presents the first high-precision, comprehensive radial-velocity (RV) survey of the cluster,
thereby providing the first census of the hard-binary population, measurements of the internal stellar 
kinematics, and improved membership information.

The recent photometric studies by \citet{von98} and \citet{sar99} used UBVRI CCD photometry to derive 
a reddening of $E(B-V) = 0.09$, an apparent distance modulus of $(m-M)_{V} = 11.44$ (1.9 kpc), and an age of 7 Gyr.  
The most recent PM study by \citet{pla03} (hereafter, P03) suggests that NGC 188 contains $\sim$1500
cluster members (membership probability $\geq$ 10\%) down to a limiting magnitude of $V$=21 within 17 pc in projection.  
We have used the P03 study to form our current cluster sample (described in Section~\ref{sam}). 

This RV analysis of NGC 188 is part of the  WIYN Open Cluster Study \citep[WOCS;][]{mat00}.
We calculate RV membership probabilities for \ttRV~solar-type stars with $\geq$3 radial-velocity measurements (Section~\ref{mem}), finding
\SM~single cluster members and \tvm~likely member velocity-variable stars (Section~\ref{var}). 
(In the following, we use the term ``single'' to identify stars with no significant RV variation.
Certainly, many of these stars are also binaries, although generally with lower total mass than the binaries
identified in this study.  We assume that the velocity-variables detected in this study are binary- or
multiple-star systems.) 
Combined with available PM data, our precise three-dimensional (3D) kinematic memberships define the 
most accurate sample of solar-type members of NGC 188 to date.   
We use this member sample to construct a cleaned color-magnitude diagram of the cluster (Section~\ref{cmd}) 
from which we identify a blue-straggler (BS) population as well as six intriguing 
stars that lie far from a standard single-star isochrone.  Finally, we investigate the  
spatial distribution of cluster member populations (Section~\ref{spat}) and the cluster's velocity 
dispersion (Section~\ref{disp}).  Following papers will provide 
orbital solutions for most of the velocity variables, likely all of which will be hard binary stars that 
dynamically power the cluster, as well as detailed comparisons with $N$-body models \citep[e.g.,][]{hur05}.

\section{Stellar Sample} \label{sam}

\subsection{WIYN Stellar Sample}

\subsubsection{Chronology}

Our ongoing WIYN\footnote{The WIYN Observatory is a joint facility 
of the University of Wisconsin-Madison, Indiana University, Yale University, and the National Optical 
Astronomy Observatories.} RV survey of NGC 188 began in 1996.  The following section describes the development of our stellar sample
over the course of these eleven years.  Our observational history as well as current strategy have been dependent on the available 
position, photometry and membership measurements.  

Initially, we composed our stellar sample from the \citet{din96} PM study, which
contains photometry and PMs ($\sigma$ = 0.5 mas yr$^{-1}$) for 1127 stars covering a 50\arcmin~x 60\arcmin~area within the 
limiting magnitudes of $B = 16.2$ and $V\approx 16.5$.   
\citet{din96} found 360 cluster members in their sample 
(probability $\geq$ 70\%).  These cluster members were prioritized in our observations (see Section~\ref{obs}).

Later, a more extensive PM study was performed by P03, complete to $V=21$ and covering 
$0.75\degr^{2}$, centered on the J2000.0 coordinates: 
$\alpha=0^h 47^m 12\fs5$ and $\delta=+85\degr 14\arcmin 49\arcsec$.  
The PMs have a precision down to 0.15 mas yr$^{-1}$ per coordinate, and the coordinates are accurate to 
5-10 mas on the International Celestial Reference System (ICRS).  
Of the 7812 objects that were detected, P03 found 1490 PM members (probability $\geq$ 10\%) across all 
magnitudes in the study.
Thus, in 2003, we chose to update our data set to include the additional stars (both PM members and non-members) 
from this study that lie within our 
magnitude limits, and to use the more accurate coordinates.  Our complete stellar sample includes 1498 stars.

\subsubsection{Sample}

The Hydra Multi-Object Spectrograph (MOS) on the WIYN 3.5m telescope has an effective dynamic range of approximately four 
magnitudes and a 1\degr~field of view.  Fortuitously, the P03 study also covers 1\degr~in 
spatial extent; we therefore did not make any spatial cut in our sample.  
We have set our limiting magnitude range to \magn, covering a large portion of the giant branch through the upper main sequence, 
including the cluster turnoff and BS stars.  At $V$ = 16.5, the faint end of our observations extends to $\sim$1.5 
magnitudes below the cluster turnoff (e.g., Figure~\ref{CMDs}). 
Our observing setup allows us to consistently determine RVs for stars whose $(\bv) \gtrsim$ 0.4; earlier type 
stars are often rapidly rotating and have fewer absorption lines in our wavelength range.  Since our sample 
includes only 30 stars blueward of this, we also observed them so as to include those that are narrow-lined.     
Currently, our stellar sample contains stars whose $(\bv)$ color lies within the range of $0.15\leq(\bv)\leq1.8$.  
After limiting the P03 study by magnitude, 
we have a total stellar sample of 1498 stars, comprising a complete 
sample of stars within our magnitude range and within 0.5\degr~(17 pc at a distance of 1.9 kpc) from the cluster center.

\subsection{DAO Stellar Sample}

The DAO data set contains multiple RV measurements of 77 stars within the 
magnitudes of 10.8$\leq$V$\leq$16 covering dates from December 1973 through November 1996.  
Observations prior to 1980 were taken by Roger Griffin and James Gunn at the 
Palomar 5m telescope \citep[e.g.,][]{gri74}; later observations were made by Robert McClure, Hugh Harris and Jim Hesser
at the Dominion Astrophysical Observatory (DAO) \citep[e.g.,][]{fle82,mcc85}.  All RV measurements
were then converted onto the DAO Radial-Velocity Spectrometer (RVS) system.

\section{Observations} \label{obs}

\subsection{WIYN}

\subsubsection{The Telescope and Instrument} \label{tel}

We have used the Hydra Multi-Object Spectrograph (MOS) on the WIYN 3.5m telescope to 
conduct our ongoing WOCS RV survey of NGC 188 from 1996 to the present.  
The telescope combines a large aperture and wide field of view, and provides excellent image 
quality (0.8 arcsecond median seeing). 
The MOS is capable of obtaining simultaneous spectra of up to $\sim$80 stars\footnote{\footnotesize 
There are currently 81 out of the 96 total usable fibers on Hydra, the remainder having been damaged 
over the course of the instrument's lifetime.} over the 1\degr~field 
during each configuration (or \textit{pointing}). The Hydra fiber positioner executes a 
fiber setup in $\sim$20 minutes with a precision of 0.2$\arcsec$. The isolated fiber-fed bench spectrograph 
is of conventional design with an all-transmission camera to avoid central obstruction of the filled beam. 
We use the blue fibers for our observations, each fiber having a 3.1$\arcsec$ aperture and an
effective spectral window of 300-700 nm.  The majority of our observations are 
centered on 513 nm and extend 25 nm in either direction, covering a rich array of narrow 
absorption lines and the Mg B triplet.  We utilize the X14 filter along 
with the Echelle grating at 11th order, obtaining a resolution of approximately 20000 and a
dispersion of 0.13 \AA/pixel on the CCD.  A portion of our observations are centered on 640 nm 
(also extending 25 nm in either direction).  The detector is a thinned 2048 x 2048 Tektronics CCD with 
24$\mu$m pixels and a typical resolution element of $\sim$ 2.5 pixels ($\sim$ 19 \kms). 

\subsubsection{Observing Procedure}

During each pointing we may utilize $\sim$80 fibers to measure simultaneous spectra.  We 
will generally use $\sim$10 fibers to measure the sky, and place the remaining $\sim$70 fibers on stars.  
Therefore, developing a strategy for the placement of these science fibers 
is critical in order to optimize limited observing time.  This section discusses our chosen 
strategy for fiber placement as well as our general observing routine.

Prior to creating our ``pointing files'' for a given observing run, we perform a detailed prioritization of the 
stars in our sample based on our available data.  
In general, we place our discovered velocity-variable stars at the highest priority, followed by 
PM members, PM non-members, and finally, those stars that we consider finished.  
Based on Monte Carlo simulations, we require 
3 epochs of RV measurements over at least a year to ensure 95\% confidence that a star is either constant or 
variable in velocity \citep[i.e., a single star or a binary- or multiple-star system; ][]{mat83,gel12}.  With three observations in hand we either derive secure 
membership probabilities for single stars and place them at the lowest priority (considering them finished), or move binaries 
to the highest priority in order to expedite orbital solutions.

To maximize progress towards orbital solutions, the short-period binaries are the highest priority 
for observation each night. Longer period binaries are the next priority for 1-2 observations per run. 
All other fibers are placed on stars for completion of the survey of the entire cluster.
\footnote{ \footnotesize These remaining fibers are placed based on the 
following ranking: ``candidate binaries'' (twice observed stars whose measured velocities vary by $>$ 1.5 \kms), followed by once-, twice-, and un-observed 
PM members, and finally once-, twice, and un-observed PM non-members.  
Each section is then sorted by distance from the cluster center, to maximize fiber density in the crowded central
field.}

During a given observing run, we produce a ``faint master'' and a ``bright master'' which will
then be used to make the initial pointing files.  The faint master contains all of the stars 
from our sample while the bright master contains only stars within the range of 
12.5$\leq$V$\leq$15.5 (777 object stars) to be used in the case where light cloud cover would 
likely prevent us from reliably deriving velocities for the fainter stars.  In general, we use a total 
of two hours of integration time for the faint pointings split into three 2400-sec 
integrations.  For bright pointings, we take one hour of total integration time split into 
three 1200-sec integrations.  By splitting up our integrations, we increase our ability 
to remove cosmic rays that may appear in our spectra.  Calibration is achieved through one 200-sec 
flat field and two 300-sec ThAr (for the 513 nm range) or CuAr (for the 640 nm range)
emission lamp spectra for each pointing.  
Currently, we have observed NGC 188 using WIYN over 53 different observing runs totalling 119 separate pointings.         

\subsection{DAO}

The DAO RVs were measured using the DAO RVS on the 1.2-meter telescope.  
This instrument observes 43 nm of the spectrum centered
at 455 nm with a resolution 0.2 \AA, scanning a mask rapidly
across the stellar absorption lines.  The properties of the instrument and the observing 
procedures can be found in \citet{fle82} and \citet{mcc85}.

\section{Data Reduction} 

\subsection{WIYN} \label{Wreduc}

The ultimate goal of our data reduction routine is to take our CCD image (the light from the 
$\sim$80 fibers), extract the $\sim$70 science spectra from this image, and obtain $\sim$70 
heliocentric RVs.  We describe our data reduction procedure in some 
detail here as the basic reference for subsequent RV publications from this project.  

Initially, we come away from the telescope with raw FITS images of the dispersed light from our science and 
calibration integrations.  We generally take two ThAr (or CuAr) calibration 
images per pointing, one before the first integration and one after the last integration, in order to account for any potential 
wavelength shift that may occur during the one or two hours of total integration.  
We also take one dome flat per pointing.  

All image processing is done 
within the IRAF\footnote{\footnotesize IRAF is distributed by the National Optical Astronomy 
Observatories, which are operated by the Association of Universities for Research in Astronomy, Inc., 
under cooperative agreement with the National Science Foundation.} 
data reduction environment \citep{tod93}.  

\paragraph{Bias Subtraction:} 
The WIYN MOS CCD contains an overscan strip of 32 x 2048 pixels, containing only 
the readout bias level.  For each row, the 32 columns are averaged. A cubic-spline function is fit to this averaged 
bias column and is subtracted from each column of the science image.  

\paragraph{Spectra Extraction:}  
First, we trace the apertures along the image; dome flats are used 
to perform this trace as the fibers are well illuminated.  We fit a Legendre polynomial function to the position of each spectrum
on the chip.  These fits are then saved, to be used later in extracting each science spectrum aperture. 
Second, we average the flat-field intensity across all fibers, fit a cubic-spline function to this single flat-field spectrum, 
and, again, save this fit for use in flat fielding the science spectra.  
Finally, the dome flat is used to correct for fiber-to-fiber throughput variations.  
Each extracted flat field aperture is averaged to a single number which is then multiplied by the average flat field intensity function
so that when the science apertures are flat fielded later, they will be throughput corrected simultaneously.  

We then determine a pixel to wavelength map using the ThAr (or CuAr) calibration spectra, first using the long (900-sec)
exposure taken during the day, and then the shorter (300-sec) exposures related to the science images.  
We use the long calibration image for our initial map, as the extracted high signal-to-noise spectra 
yield more precise results.  After extracting the spectra from the long calibration image, we determine 
a dispersion solution for each aperture by fitting a cubic-spline function to the ThAr (or CuAr) emission line 
wavelengths and pixel centroids.  We then transfer these solutions as initial guesses onto the extracted 
short calibration spectra associated with our science spectra.
Our typical rms residuals around the emission line wavelengths for a fit to a long calibration spectrum vary 
from $\sim$0.006-0.009 \AA, and increase slightly for the short calibration exposures.  

Finally, the science spectra are extracted, flat fielded, throughput corrected, and dispersion 
corrected.  The result is a multi-spectrum FITS image containing all science spectra for a 
given integration.  The preceding steps are performed for each of a set of integrations for a given field using the 
associated flat fields and calibration images.

\paragraph{Sky Subtraction:}
We generally do not observe during dark time and therefore often obtain some scattered moonlight 
in our object spectra.  In order to avoid the potential confusion due to a second 
spectrum of different velocity, we must remove the solar absorption lines from our spectra.  
This is achieved through the use of our $\sim$10 sky fibers.  We inspect each sky spectrum and then use all 
acceptable sky spectra from our pointing to create one averaged sky spectrum per integration. (Inclusion of 
a noisy, low quality or accidental stellar spectrum will degrade our science spectra when sky subtracted.) 
During the averaging, the sky spectra are run through an averaged sigma clipping algorithm in order to remove cosmic rays.  
Finally, this combined sky spectrum is subtracted from each object spectrum, and the procedure is 
repeated for all integrations. 

At this point, we generally have three extracted and sky subtracted sets of spectra for each pointing
(i.e., one set of spectra for each integration of a given pointing), which we combine to produce one set of 
spectra per pointing.  We perform a median filter combination to remove any 
cosmic ray contamination from our object spectra.  

\paragraph{Cross Correlation:}
We then calculate RVs from these combined object spectra using the IRAF 
task \textit{fxcor}.  This task uses the fourier transforms of two spectra (generally one science and one template spectrum)
to quickly compute the 
cross-correlation function (CCF) between them, and hence derive their velocity difference.  
We fit a Gaussian function to the CCF peak and find the centroid, obtaining the velocity offset 
of the object spectrum from the template.  We allow IRAF to correct this velocity for the template 
offset and the motion of the earth around the sun to return a heliocentric velocity.  These 
heliocentric velocities are the final product of our data reduction routine.

Since we are observing solar-type stars, we have chosen to correlate each object spectrum against a 
high signal-to-noise solar spectrum obtained from a daytime sky observation.  
This solar template spectrum was obtained using the same 
observing setup as used for our science spectra.  Currently, the same template is used to reduce all of our 
WIYN observations.  However, certainly not all of our stars have a G2V spectral type.  Therefore 
we have investigated for a potential offset in velocity resulting from a template mismatch
by binning our measured single member star RVs as a function of $(\bv)$~color. 
There is no evidence for a systematic offset to the level of 0.1 \kms. We are therefore confident that our choice of a G star 
template has not affected the accuracy of our RV measurements.

\subsection{DAO}

Information regarding the DAO RVS data reduction routine can be found in \citet{fle82} and \citet{mcc85}.

\section{WIYN Data Quality Analysis} \label{dat}

Here we present a thorough analysis of our WIYN RV measurement precision and the various sources of dispersion in our 
measurements. In Section~\ref{qual}, we briefly discuss our quality control
procedures. In Section~\ref{prec} we derive our measurement precision through the use of the $\chi^2$ function.
Finally, we provide a detailed empirical study of our error budget in Section~\ref{erbud}.

\subsection{Quality of Measurement} \label{qual}

\begin{figure}[!t]
\plotone{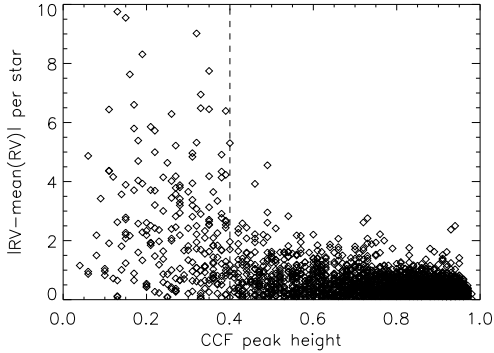}
\caption{\footnotesize RV offset as a function of the CCF peak height. 
Each data point represents the offset of an RV measurement
of a star from the mean RV of that star,
 plotted against the measurement's CCF peak height. Only single stars were used.  
The mean RV for each star was calculated 
using only measurements whose CCF peak heights were $>$ 0.4.  Individual measurements with CCF peak heights $>$ 0.4 show 
small offsets from the average, while those with CCF peak heights $<$ 0.4 show a much larger scatter.}
\label{ccfcut}
\end{figure}

Generally the highest signal-to-noise spectra result in 
the highest CCF peak height values. We use the peak height of the CCF fit as the measure of the 
quality of a given measurement in the WIYN data set.  Only measurements whose 
fits lie above a certain cutoff value in CCF peak height are deemed reliable. In order to determine this 
cutoff value, we first determine the offset of each RV measurement
of a star from the mean RV of that star, using only single stars. We then plot these offsets, for all
RV measurements of all single stars, against CCF peak height (Figure~\ref{ccfcut}). 

It is clear from Figure~\ref{ccfcut} that above a CCF peak height of $\sim$0.4 the RV offsets are 
quite small while those measurements whose CCF peak heights are lower than 0.4 show a very large 
scatter.  We therefore adopt a CCF peak height cutoff value of 0.4, and present
here only measurements of this quality.

In addition to our CCF peak height quality criterion, we examine the distribution of RVs for each individual star and 
visually inspect any measurements that are outliers in the distribution.  Occasionally we remove measurements whose
CCF, though having a peak height above 0.4, clearly provides a spurious
measurement (e.g., inadequate sky subtraction).  

\subsection{Precision} \label{prec}

In order to determine the precision of our WIYN RV measurements, we fit a $\chi^{2}$ distribution to 
our empirical distribution of RV standard deviations (following \citet{kir63}).  The 
theoretical $\chi^{2}$ curve models the distribution of errors on measurements of single stars.  This function depends on both the precision of all measurements as well as the 
number of degrees of freedom (i.e., one less than the number of measurements).  For consistency across all stars, many of which have only three measurements, here we have used only the first three measurements of each star to calculate the standard
deviation of its measurements\footnote{We have investigated the use of the \citet{kee62} small-number statistical 
formulae to calculate these standard deviation values, and find that these formulae suggest only a small 
($\sim$10\% for three measurements) correction to standard deviations measured using the traditional formula.
Therefore we have decided to use the traditional formula to calculate our standard deviation values, and 
note that we derive the same global precision on our measurements using either method.}.  
In order to derive this precision value, we fit a $\chi^{2}$ function with 2 degrees of freedom 
to the distribution of the RV standard deviations:   
\begin{equation}
\chi^2(\sigma_{obs})  =  \left(\frac{2}{\sigma_i^{2}}\right) \left(\sigma_{obs}\right) \exp\left({-\frac{\sigma_{obs}^{2}}{\sigma_i^{2}}}\right)
\end{equation}
Here, $\sigma_{obs}$ denotes the standard deviation of the RV measurements for a star, and $\sigma_i$
represents the overall precision of our measurements.  

\begin{figure}[!t]
\plotone{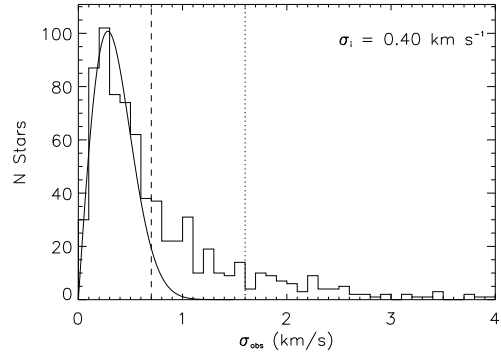}
\caption{\footnotesize Histogram of the RV standard deviations of the first three WIYN RV measurements 
for all stars observed $\geq$ 3 times. Also shown is the best-fit $\chi^{2}$ distribution function.  The fit yields a 
measurement precision of 0.40 \kms.  The dashed line shows our maximum $\sigma_{obs}$ of 0.7 \kms; the measurements
at $\sigma_{obs} >$ 0.7 \kms~were excluded from the $\chi^2$ curve fit due to contamination from velocity variables.  The dotted line shows our cutoff value of 
4 times $\sigma_i$, above which we consider a star to be a secure velocity variable.} 
\label{3all.eoveri}
\end{figure}

The empirical distribution (Figure~\ref{3all.eoveri}) represents the combination of two populations: 
the single stars (with lower $\sigma_{obs}$), and the velocity-variable stars 
(with higher $\sigma_{obs}$).  
The $\chi^{2}$ function is designed to fit only the single stars.  For this reason, we chose a maximum 
$\sigma_{obs}$ for our fit, above which the population is dominated by velocity-variable stars.  
We used a standard least-squares determination, varying the value $\sigma_i$ as well as a cutoff value in $\sigma_{obs}$, to fit the $\chi^2$ function to our data.  
A maximum $\sigma_{obs}$ cutoff of 0.7 \kms~(shown by the dashed line in Figure~\ref{3all.eoveri}) and $\sigma_i$ = 0.4 \kms~
produced the smallest residuals and hence the best fit to the peak at lower $\sigma_{obs}$.
Figure~\ref{3all.eoveri} shows the RV standard deviation histogram of all stars with $\geq3$ WIYN measurements along with the best 
fit $\chi^2$ distribution function.  This analysis yields a nominal precision on our WIYN data of 
\boldmath $\sigma_i$ \unboldmath \textbf{= \pr}.

Note that towards higher standard deviations, the data are populated almost entirely by binary stars.  
This is reflected in the departure of the data from the theoretical $\chi^{2}$ curve.  We can 
therefore use this plot to facilitate our binary identification routine by assuming that if a given star has 
a $\sigma_{obs}$ that is higher than a certain value, the star is a velocity 
variable.  By inspection of Figure~\ref{3all.eoveri} we note that the theoretical curve drops 
to very near zero at $>$ 3 times $\sigma_i$.   We therefore  define our velocity-variable stars as those stars whose 
RV standard deviations (by convention, we use \textit{e} in place of $\sigma_{obs}$) are 
greater than 4 times our quoted precision (by convention, we use \textit{i} in place of $\sigma_i$), or \eoi~$>$~4.
We plot this value of \eoi~= 4 as the dotted line in Figure~\ref{3all.eoveri}, and we discuss our velocity variables in 
further detail in Section~\ref{var}.  

\begin{figure}[!t]
\plotone{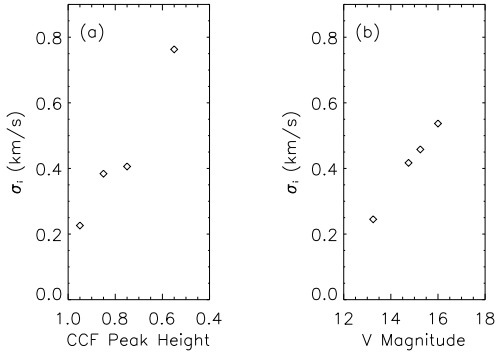}
\caption{\footnotesize Plots of our measurement precision as a function of 
CCF peak height (a) and apparent $V$ magnitude (b).  The $\sigma_i$ values are derived via the 
$\chi^2$-fit analysis discussed above.} 
\label{V.ccf.eoveri}
\end{figure}

Although we have calculated a single value for our WIYN precision 
the dispersion of our measurements is affected by photon error, and is thus a function of $V$ magnitude.    
Figure~\ref{V.ccf.eoveri} shows $\chi^2$ calculated values of $\sigma_i$
as functions of CCF peak height and $V$ magnitude.  For 
fit quality, our precision ranges from 0.22 \kms~for the highest quality RVs and degrades to 
0.80 \kms~towards our CCF peak height cutoff value of 0.4.  
We see a similar range when looking at the precision as a function of $V$ magnitude; 
the brightest stars are measured to a precision 
of 0.25 \kms~while the measurement precision on our faintest stars is reduced to 0.55 \kms.  

The range in $\chi^2$-fit values for our precision agrees very well with that derived from our error 
budget analysis in Section~\ref{erbud}.  We are therefore confident in quoting our overall precision 
to be $\sigma_i$ = \pr, which can then be used to identify our binary stars and to
evaluate the velocity dispersion of the cluster.    

\subsection{Empirical Error Budget Analysis} \label{erbud}

In this section, we present a thorough empirical analysis of the error sources contributing to our measurement precision; 
a preliminary report can be found in \citet{mei01}. 
Our internal measurement errors have both systematic and random components that include
fiber-to-fiber variations ($\sigma_{f-f}$), photon error ($\sigma_{ph}(V)$), and pointing-to-pointing variations ($\sigma_{p-p}$), 
each of which we discuss and quantify here.

\subsubsection{Systematic Fiber-to-Fiber Variations ($\sigma_{f-f}$):}

Sky flats demonstrate the existence of fiber-to-fiber variations of unknown origin.  
During our eleven years of monitoring the stars in NGC 188, 
different configurations of the Hydra MOS result in the same star being measured through 
different fibers. Thus fiber-to-fiber variations represent a source of internal error on the set of 
measurements for a star .  Fortunately this error is systematic and can be corrected.
 
\begin{figure}[!t]
\plotone{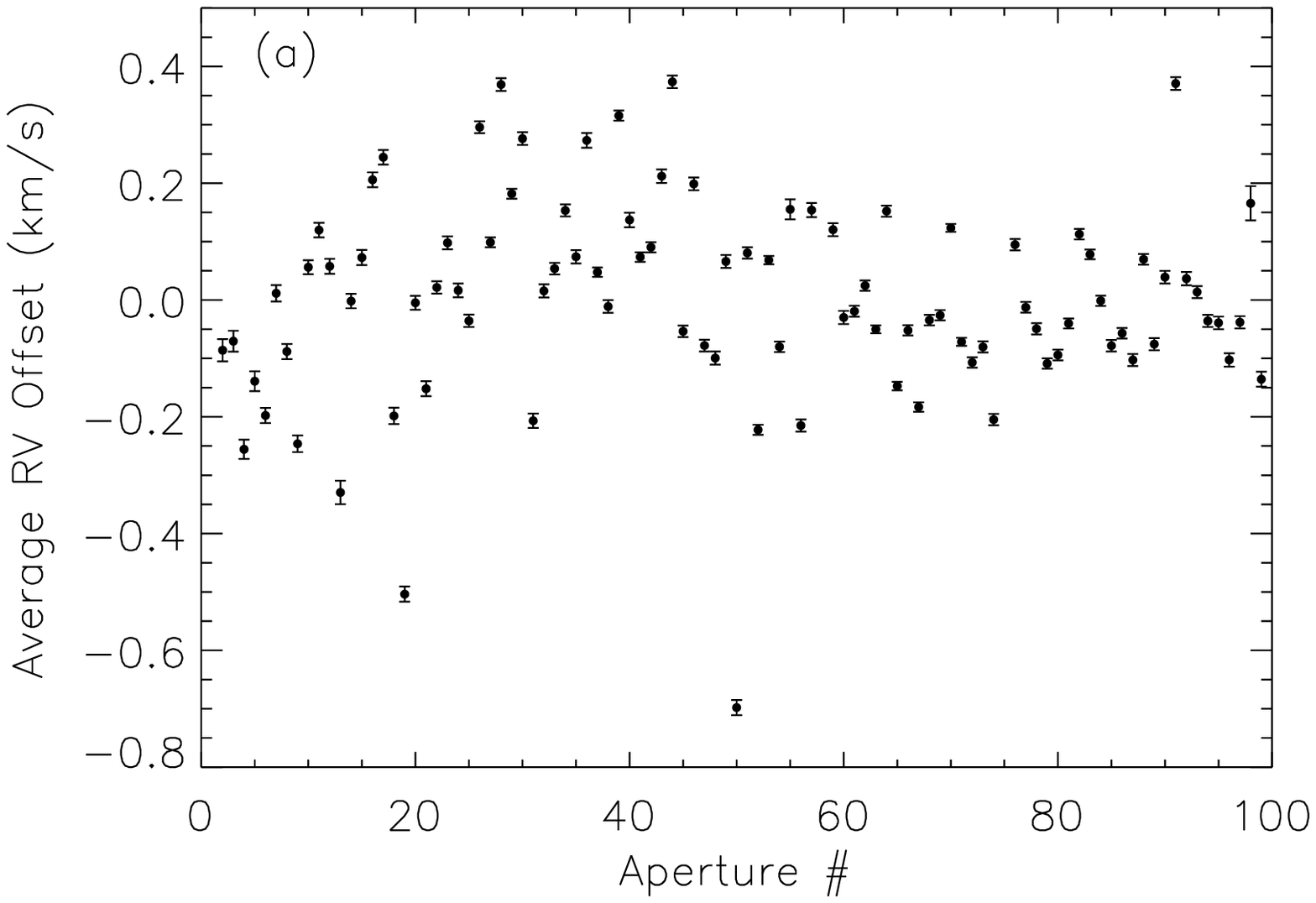}
\plotone{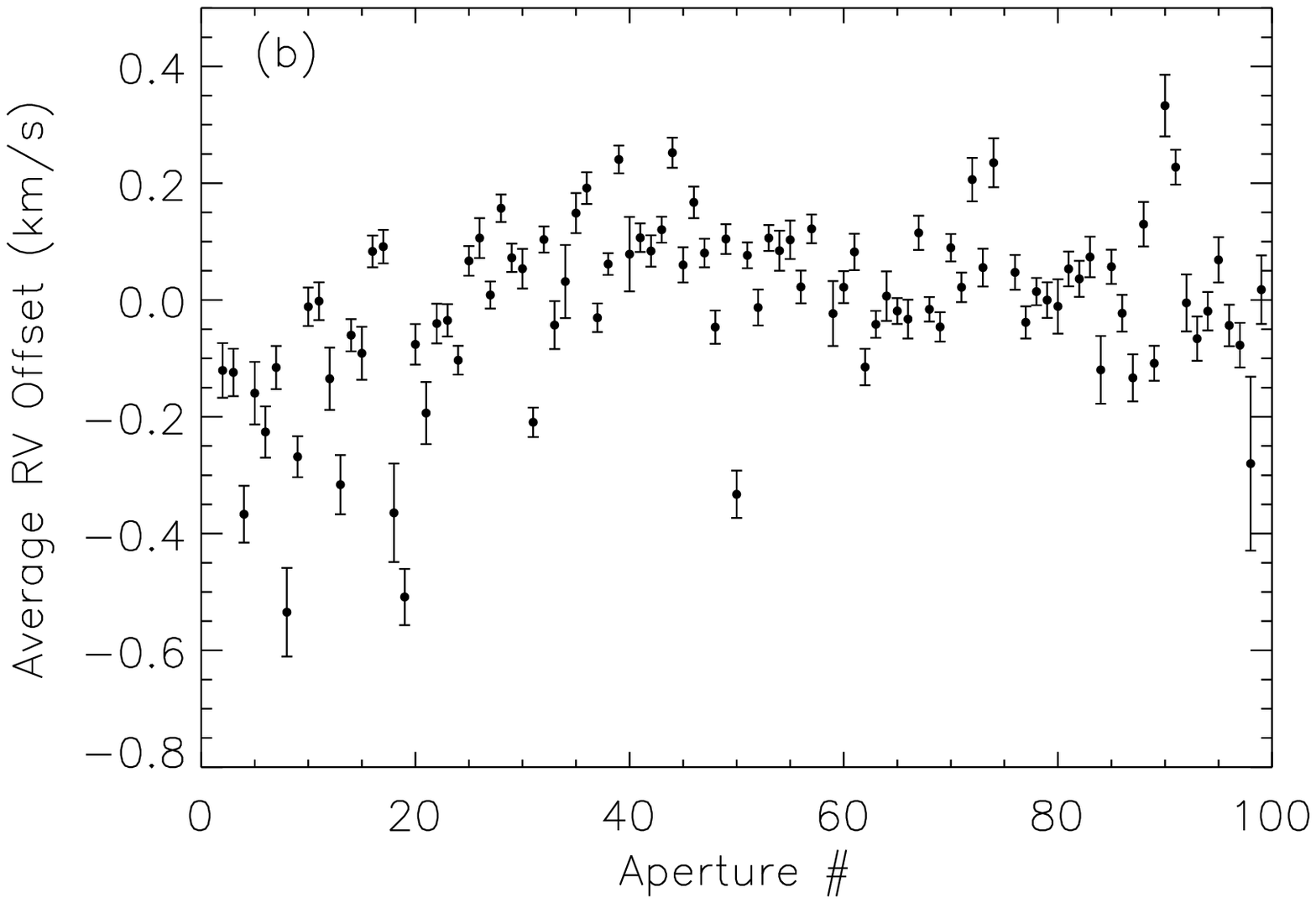}
\caption{\footnotesize RV offsets for each Hydra MOS fiber.  These offsets were 
calculated using sky flats spanning $\sim$10 years.  Plot (a) shows fiber offsets for the 
513 nm range, and (b) shows offsets for the 640 nm range.  The error bars 
represent the RMS error on the mean of each data point. (Errors are larger for the 640 nm
region simply because we have fewer observations at this wavelength range.)}
\label{fiber}
\end{figure}

During each observing run, we take at least one sky flat as a calibration measure, with all fibers arranged in a circle in the focal plane. When corrected for the Earth's rotational 
and orbital motions, all fibers should provide radial velocities of 0 \kms. 
In reality, it has become clear that each fiber has a unique and consistent offset of $<$ 1 \kms. Specifically, for each sky flat we derive a mean
velocity across all of the fibers, and then measure the velocity offset from this mean for each fiber.
We take the means of these velocity offsets over $\sim$10 years of observing NGC 188 for each of the two observed wavelength 
ranges, and the results are plotted in Figure~\ref{fiber}.  The offsets for the 513 nm region range from 
0.00 \kms~to 0.70 \kms.  The mean of the standard deviations on each mean offset measurement is 0.13 \kms.  
Analysis of the 640 nm range results in similar values.  We assume that the derived offset for a given fiber is independent 
of the position of the fiber on the focal plane.  Our RV measurements are corrected for these offsets on a fiber-by-fiber 
basis.
 
\subsubsection{Random Photon Errors ($\sigma_{ph}(V)$):}

\begin{figure}[!t]
\plotone{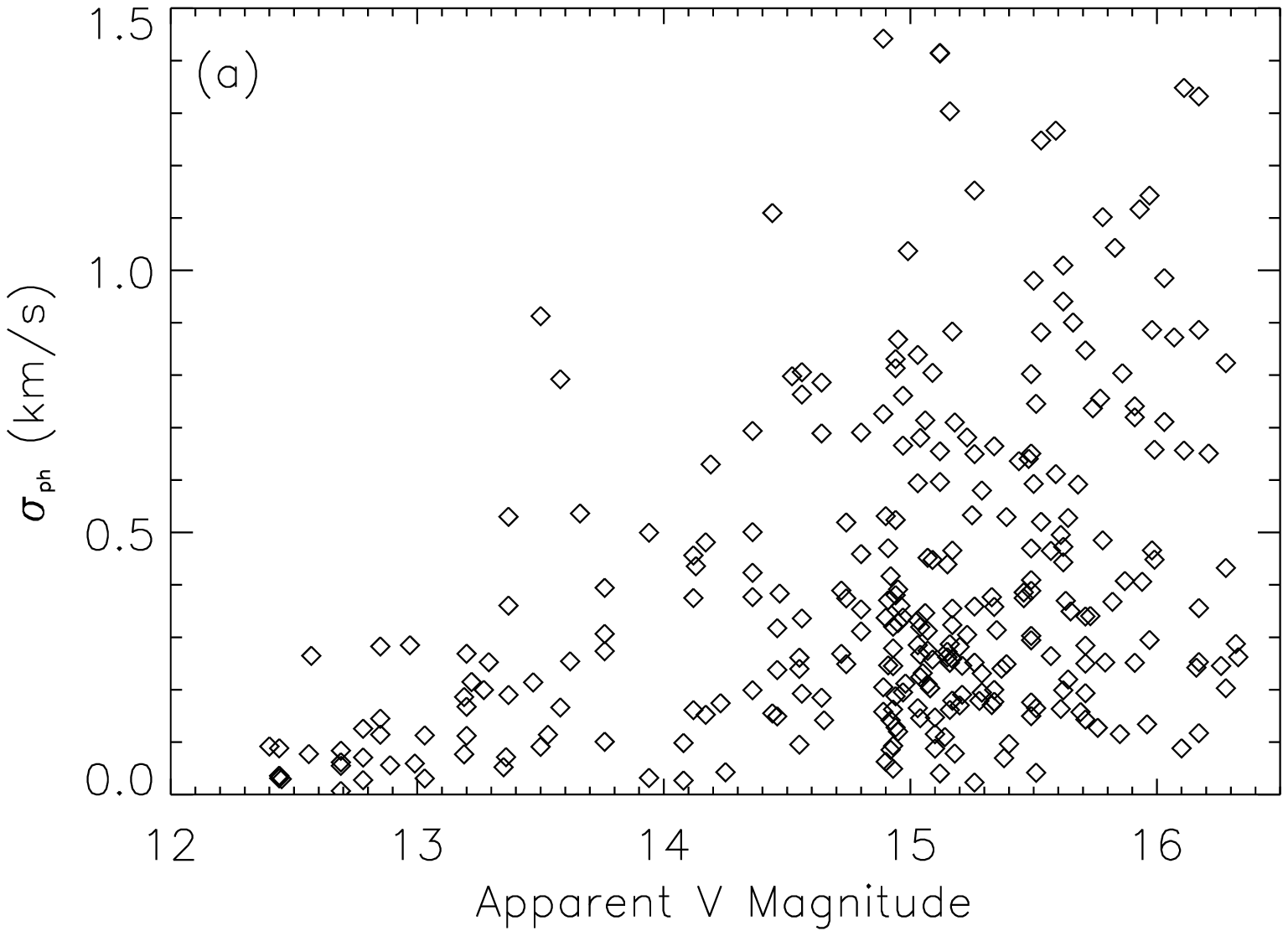}
\plotone{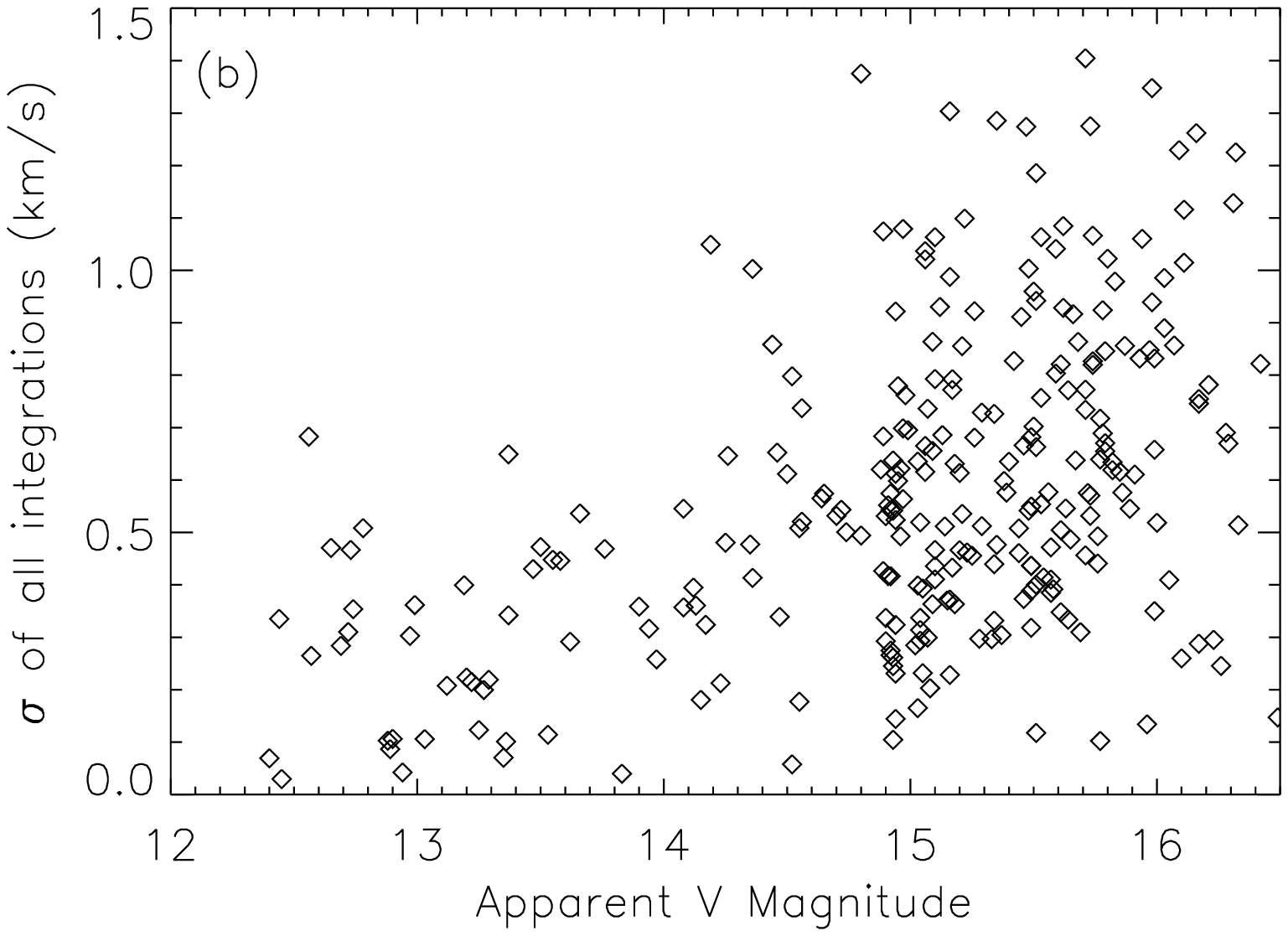}
\caption{\footnotesize (a) Standard deviations of three sequential 2400s RV measurements as a function of apparent $V$ 
magnitude.  These standard deviations derive only from photon error.
(b) Standard deviations of all RV measurements for single stars as a function of apparent $V$ 
magnitude.}
\label{photer}
\end{figure}

As described in Section~\ref{obs}, 
each RV measurement derives from three combined sequential integrations obtained in the same fiber configuration.  Since these three integrations are 
almost contemporaneous and are made through the same fibers for any given star, the standard deviations of the three velocities measured from each of the three integrations must arise from photon error alone.
Figure~\ref{photer}(a) shows the distribution of such standard deviations as a function of $V$ magnitude, and provides a direct measurement of our photon error. It is clear that the photon error is a function of $V$ magnitude, as expected. (Here we have only included 2400-second integrations.)  Indeed, if we bin the data shown in Figure~\ref{photer}(a) by intensity ($I$), the 
trend is well fit by $I^{-1/2}$.

\subsubsection{Pointing-to-Pointing Variations ($\sigma_{p-p}$):}

In Figure~\ref{photer}(b), we plot the standard deviations of the RV measurements for each single star as a function of $V$ magnitude.  
For $V$ $\gtrsim$ 14, the morphologies of Figures~\ref{photer}(a) and 
\ref{photer}(b) are 
nearly identical.  For these fainter stars, the dispersion on our measurements is dominated by photon error.   
However, at $V$ $\lesssim$ 14, the values in Figure~\ref{photer}(a) drop below those in Figure~\ref{photer}(b).  
Here, the photon error is lower than our empirical precision.   
Indeed, at $V$ $<$ 14, we observe a constant dispersion with an average value of $\sim$0.3 
\kms~while our average photon error in Figure~\ref{photer}(a) is only $\sim$0.2 \kms.  
We must therefore identify a second component to our random error.  We conjecture that this component is due to 
systematic offsets in each pointing that are randomly distributed in time.  We label this component the 
pointing-to-pointing variation: $\sigma_{p-p}$.

Thus, we define our internal sources of random error as:
\begin{equation} \label{ine}
\sigma_i(V)^2 = \sigma_{ph}(V)^2 + \sigma_{p-p}^2
\end{equation}     

Because we do not have multiple observations of identical pointing configurations, we need to determine $\sigma_{p-p}$ via a 
statistical approach. We begin by defining the quantity $\sigma_{mp}$, the dispersion of the mean RVs of all pointings.  Specifically, 
for each pointing we take the mean RV of the single cluster members, which we call the pointing velocity. Then $\sigma_{mp}$ is the standard 
deviation of all of these pointing velocities. 

Note that $\sigma_{mp}$ is not simply $\sigma_{p-p}$, because each pointing velocity derives from a different sample of single cluster members.
The dispersion among pointing velocities is thus a consequence of the pointing-to-pointing variations, the photon errors, and the true 
motions of the stars.  We digress here briefly to discuss the last.

Stars in a gravitationally bound cluster will orbit the center of mass of the system. The orbital motions around the cluster center-of-mass 
is typically characterized by a velocity dispersion ($\sigma_c$). In addition, orbital motions introduced by a binary companion will offset 
the individual RV measurements of a given star from the binary system's center-of-mass motion. When a binary orbital solution is available 
the true center-of-mass motion is known. However, there are certainly binaries that we cannot detect spectroscopically 
(e.g., long-period binaries), and are thus included as single (constant-velocity) stars.  These undetected binaries add to the dispersion on 
our measurements.  We characterize the distribution of undetected binary motions as a Gaussian with dispersion
$\sigma_b$. Therefore, we call the combination of the cluster motions and undetected binary motions the combined velocity dispersion:
\begin{equation}
\sigma_{cb}^2 = \sigma_c^2 + \sigma_b^2
\end{equation}     

Thus, we define the 
measured pointing variation as:
\begin{equation} \label{mp}
\sigma_{mp}^2 = \frac{[\sigma_{cb}^2 + \sigma_{ph}(V)^2 ]}{N} + \sigma_{p-p}^2
\end{equation}
where $N$ corresponds to the average number of single cluster member stars in a given pointing 
(within the desired magnitude range).

Using equation~\ref{ine}, the apparent dispersion of the RV measurements for a set of stars can be written as: 
\begin{equation} \label{scm2}
\sigma_{obs}^2 = \sigma_{cb}^2 + \sigma_i^2 = [\sigma_{cb}^2 + \sigma_{ph}(V)^2 ] + \sigma_{p-p}^2
\end{equation}
where $\sigma_{obs}$ is the standard deviation of all single stars' mean velocities (in the 
appropriate magnitude domain).

Combining equations~\ref{mp} and~\ref{scm2} allows us to eliminate 
the bracketed values and recover the pointing-to-pointing dispersion:
\begin{equation} \label{ppdisp}
\sigma_{p-p}^2=\frac{N}{N-1} (\sigma_{mp}^2 - \frac{\sigma_{obs}^2}{N})
\end{equation}

\subsubsection{Calculating the Empirical Internal Error  ($\sigma_i(V)$) :}

Since the photon error is magnitude dependent, we have chosen to divide our calculation into two magnitude ranges: 
(1) 12$\leq$V$<$14 and (2) 15$<$V$\leq$16.5.

We can calculate values for $\sigma_{ph}$ directly from the analysis that lead to Figure~\ref{photer}(a). 
Next, in domain (1) we find $\sigma_{mp}$ = 0.33 \kms~, while in domain (2) we find $\sigma_{mp}$ = 
0.36 \kms.  Note, it is just fortuitous that both of these values are so close, as the larger photon error at $V$ $>$ 15 
is offset by the larger number of stars in this magnitude range.  
Using these values, we find a $\sigma_{p-p}$ to equal 0.26 \kms~in domain (1) and 0.33 \kms~in domain (2). The closeness of these 
values leads us to suspect that the origin of these small pointing-to-pointing variations is not magnitude dependent.
We proceed to combine our error budgets to compute our internal error in our two magnitude domains. 
Table~\ref{ervals} summarizes our results for the error budgets and the internal error as a function of magnitude.

\begin{deluxetable}{l c c }
\tabletypesize{\small}
\tablewidth{0pt}
\tablecaption{Empirical Internal Errors\label{ervals}}
\tablehead{\colhead{} & \colhead{(1) 12$\leq$V$<$14} & \colhead{(2) 15$<$V$\leq$16.5}}
\startdata
$N$ (stars per pointing) & 10 & 20 \\
$\sigma_{obs}$ (\kms) & 0.70 & 0.76 \\
$\sigma_{mp}$ (\kms) & 0.33 & 0.36 \\
$\sigma_{p-p}$ (\kms) & 0.26 & 0.33 \\
$\sigma_{ph}$ (\kms) & 0.20 & 0.51 \\ \\
\boldmath $\sigma_i$ \unboldmath \textbf{(\kms)} & \textbf{0.33} & \textbf{0.61} \\
\enddata
\end{deluxetable}

Thus we find our internal error in region (1) to be $\sigma_i$ = 0.33 \kms~and in region (2) to be $\sigma_i$ = 0.61 
\kms.  These values match very closely to those derived in our $\chi^2$ analysis presented in Section~\ref{prec} and 
Figure~\ref{V.ccf.eoveri}.

\section{The Combined WIYN and DAO Data Set} \label{comb}

In our database, we currently have a total of \tot~RVs of \tRV~stars from our NGC 188 cluster sample.  
Of those, 544 RVs of 77 stars were measured at the DAO (or at the Palomar 5m telescope for those prior to 1980); 
the remaining data were measured at the WIYN 3.5m telescope.  
Before combining these two data sets, we searched for a potential zero-point offset and for differences in precision.

\begin{figure*}[!t]
\plotone{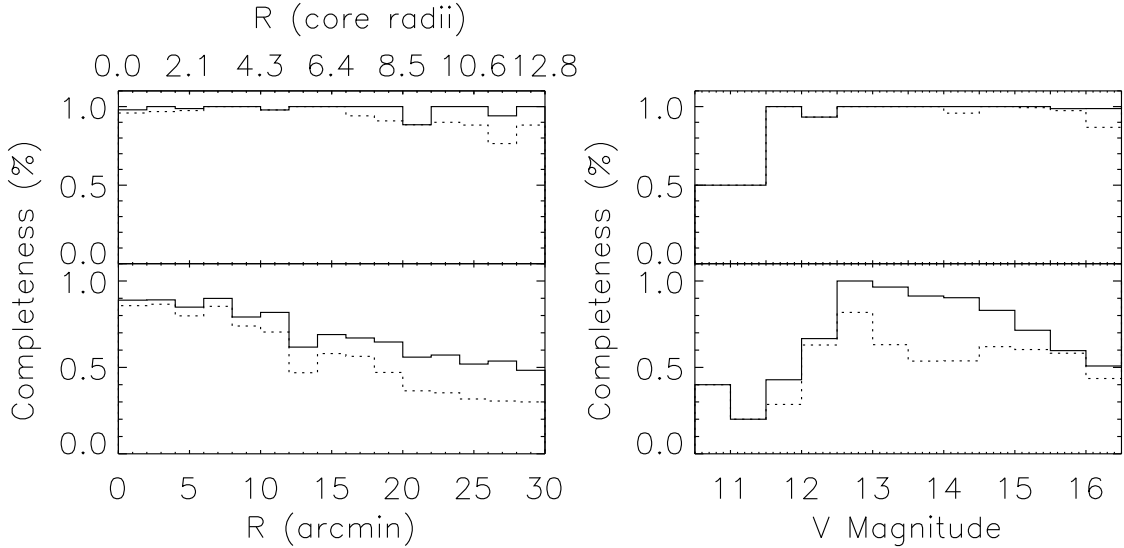}
\caption{\footnotesize Completeness histograms as a function of radius (left) and apparent $V$ magnitude (right) for our NGC 188 sample.
In the left panels, we show radius in both arcminutes and core radii, assuming a core radius of 1.3 pc \citep{bon05} at a 
distance of 1.9 kpc.
Our percentage of stars observed $\geq$1 time is plotted with the solid line, while our percentage of stars observed $\geq$3 
times is plotted with the dotted line.  The top panels show our completeness for the P03 non-zero PM membership stars,
and the bottom panels show our completeness for all stars in our sample.}
\label{compfig}
\end{figure*}

\subsection{Zero Point and Precision}

In order to check for a potential zero point offset between the two data sets, we compared 24 single
stars with $\geq$3 observations in each data set.  A star 
is defined as single if the standard deviation of its measurements is 
less than four times our measurement precision of \pr~(see sections~\ref{prec} and~\ref{var}).
We find a mean offset of 0.03 $\pm$ 0.1 \kms.  We therefore conclude that there is no measurable 
systematic RV offset between the WIYN and DAO data at this level.

We then investigated the precision of each data set, employing a $\chi^2$-fitting algorithm 
as described by \citet{kir63} and discussed in Section~\ref{prec}.  Using this method, we find 
the precision of the WIYN data set to be \pr~and of the DAO data set to be 1.0 \kms.  
The DAO RV errors found here are larger than the typical
errors of 0.4 km s$^{-1}$ normally seen with this instrument
\citep{fle82}.  We believe these increased errors are due to
photon statistics, as the stars in NGC 188 are faint enough
to require long exposure times for this relatively small
telescope.
We have therefore noted the number of WIYN and DAO 
observations for each star in our master data table.  We have also accounted for this difference in 
precision when calculating our average RVs and \eoi~measurements for stars observed at both observatories (as 
explained in Sections~\ref{mem} and Section~\ref{var}).
  
The WIYN and DAO RV measurements were thus integrated without modification
into a single data set.  We note that the combination of the data sets extends our maximum time baseline up to 
35 years for some stars.

\subsection{Completeness} \label{comp}

The basis for our stellar sample is the P03 PM study of NGC 188, which we take to be complete within 
our magnitude limits.  We note that 28 of these 1498 stars lack PM measurements.
 
As mentioned in Section~\ref{sam}, our original cluster sample was composed of the PM members from the   
\citet{din96} study, and, as such, our observations are most complete within this sample.  
We have $\geq$3 observations of 99\% of the \citet{din96} sample with PM membership probabilities $>$ 
70\%.  The Dinescu et al. PM members (P$_{PM} >$ 70\%) comprise 27\% of their entire sample. 
We have also observed 47\% of the remainder of the Dinescu et al.~sample $\geq$3 times, and 65\% at least once.

Our current stellar sample consists of 1498 stars, including 624 
with non-zero PM membership probabilities, taken from the P03 study.  
We have $\geq$3 observations of 96\% of these non-zero PM members, and at least one observation of $>$99\% of these members.  
We have also observed 27\% of the remainder of the Platais et al.~sample $\geq$3 times, and 46\% at least once.
We have at least one observation of 69\% of our entire stellar sample.

In total, we have calculated RV membership probabilities for \ttRV~stars (each having $\geq$ 3 RV measurements).    

We have plotted our completeness as a function of radius in the left panel of 
Figure~\ref{compfig} and as a function of $V$ magnitude in the right panel of Figure~\ref{compfig}.  The top plots
display our completeness in the stars with non-zero P03 PM membership probabilities.  Our observations near completeness 
in both radius and magnitude for this sample.  There is a slight bias towards the cluster center as a result of having 
prioritized these stars in our observations.  
Our completeness drops for magnitudes brighter than V=12 as we have not observed 
these stars at the WIYN 3.5m; the data for these stars comes from the DAO.  There are only 6 stars with 10.5$\leq$V$<$12 and 
non-zero P03 PM membership probabilities.  
Additionally our completeness amongst stars observed 
$\geq$3 times begins to drop at $V$ $\gtrsim$ 16.
These stars are more difficult to observe, as they require clear skies and large separations from the moon. 
The bottom plots display our completeness for our entire stellar sample.  
The drop in completeness towards larger radii is due to a similar decrease in frequency of PM members at increasing radius from the 
cluster center.  The incompleteness in $V$ $>$ 12 within our full stellar sample is again due to a decreasing frequency of PM 
members among fainter stars.

\section{Results} \label{res}

We have compiled a table of all observed stars within our survey and show the first ten lines in Table~\ref{RVtab}, while 
the entire table is provided electronically.  
For each star, we have included the WOCS (P03) ID ($ID_W$), 
\citet{san62} ID ($ID_{SA}$), \citet{din96} ID ($ID_D$), and \citet{ste04} ID ($ID_{ST}$).  
We also include the P03 RA, DEC, $V$, and $(\bv)$.
Next we show the number of WIYN ($N_W$) and DAO ($N_D$) observations, average RV ($\overline{RV}$), P03 PM membership ($P_{PM}$), 
RV membership ($P_{RV}$), \eoi~measurement, class (described in Section~\ref{mem}), and a comment.  
We use a weighted average to calculate the mean RV and \eoi~measurements.
In the comment field, we use G for giant stars, BS for blue stragglers, SB1 and SB2 for single- and 
double-lined spectroscopic binaries, respectively, and RR for rapid rotators. We define the rapid rotators as those stars 
whose CCF FWHM is $>$ 60 \kms~and are not observed to be an SB2 binary. (Our mean FWHM for non-RR stars is 35 \kms.) 
We have also included the relevant IDs 
from the photometric and X-ray studies of the cluster (with prefix ``V'' from \citet{zha02}, ``WV'' from \citet{kaf03},
``X'' from \citet{bel98}, and ``GX'' from \citet{gon05}), as well as certain short-period variable classifications from
\citet{zha02}.  
We caution the reader that the variability of cases where, for example, \eoi~$<$ 5 and only a few observations are 
available should be considered uncertain.

\begin{deluxetable*}{l l l l c c c c c c c c c c l l}
\tabletypesize{\tiny}
\tablecaption{Radial Velocity Data Table \label{RVtab}}
\tablehead{\colhead{ID$_W$} & \colhead{ID$_{SA}$} & \colhead{ID$_D$} & \colhead{ID$_{ST}$} & \colhead{RA} & \colhead{DEC} & \colhead{V} & \colhead{B-V} & \colhead{N$_W$} & \colhead{N$_D$} & \colhead{$\overline{RV}$} & \colhead{P$_{PM}$} & \colhead{P$_{RV}$} & \colhead{e/i} & \colhead{Class} & \colhead{Comment}}
\tablewidth{0pt}
\startdata
52 &  &  & 501 & 00:24:48.705 & 85:34:14.36 & 16.370 & 0.661 & 1 & 0 & 12.33 & 31 &  &  & U &  \\
91 &  &  & 857 & 00:26:49.002 & 85:27:19.11 & 16.120 & 0.789 & 5 & 0 & 23.86 & 57 & - & 6.16 & BLN & SB1 \\
96 &  & 1924 & 1158 & 00:27:33.157 & 85:34:59.40 & 13.908 & 0.706 & 13 & 0 & -29.22 & 0 & - & 39.97 & BU & SB1 \\
117 &  & 1916 & 1379 & 00:28:39.664 & 85:30:37.77 & 14.078 & 1.117 & 9 & 0 & -50.65 & 0 & 0 & 0.48 & SN &  \\
125 &  & 1938 & 973 & 00:27:09.762 & 85:29:30.67 & 15.598 & 0.787 & 3 & 0 & -84.44 &  & 0 & 2.74 & SN &  \\
136 &  & 1691 & 1229 & 00:28:19.530 & 85:28:08.38 & 14.856 & 0.928 & 3 & 0 & -94.02 & 0 & 0 & 2.61 & SN &  \\
142 &  &  & 1188 & 00:28:13.319 & 85:27:24.98 & 16.000 & 1.030 & 3 & 0 & -12.46 & 0 & 0 & 3.60 & SN &  \\
149 &  &  & 1790 & 00:29:39.550 & 85:40:28.12 & 14.156 & 0.214 & 1 & 0 & 5.11 &  &  &  & U &  \\
160 &  & 1929 & 1916 & 00:30:20.571 & 85:38:31.55 & 13.531 & 0.663 & 3 & 0 & -39.95 & 0 & 21 & 3.50 & SN &  \\
200 &  &  & 1964 & 00:30:57.379 & 85:31:14.40 & 16.111 & 0.790 & 7 & 0 & -63.76 & 43 & - & 13.17 & BLN & SB1 \\
\enddata
\tablenotetext{}{
\begin{center}
\footnotesize
\vspace{-1.5em}
(This table is available in its entirety in a machine-readable form in the online journal. A portion is shown here for guidance regarding its form
and content)
\end{center}
}
\end{deluxetable*}

\subsection{Membership} \label{mem}

\begin{figure}[!t]
\plotone{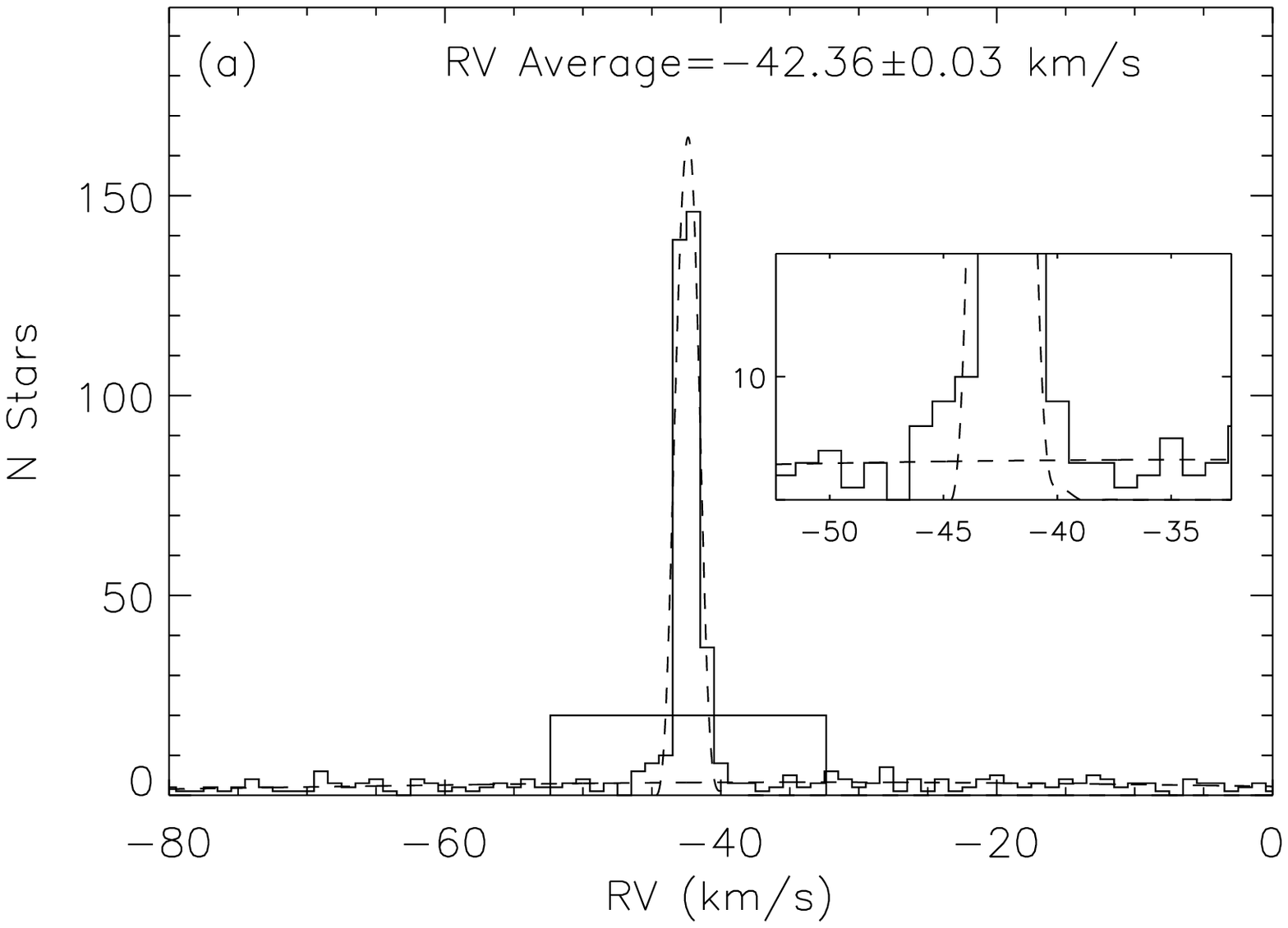}
\plotone{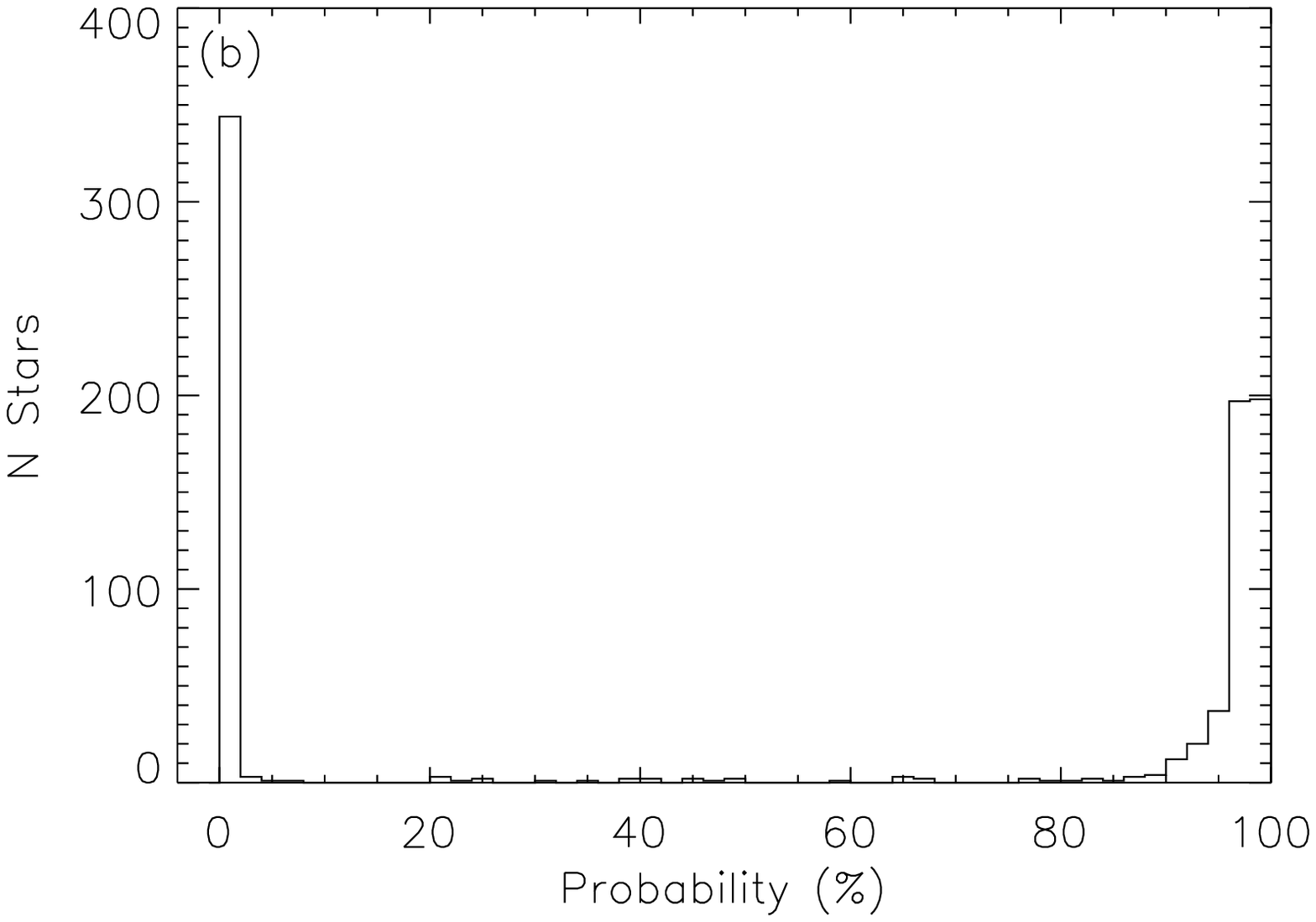}
\caption{\footnotesize (a)  RV distribution of single stars in the direction of NGC 188.  The 
cluster and field are both fit with Gaussian function which are used in calculating the 
RV membership probabilities.  The boxed region in the bottom center of the figure is expanded to the 
upper right in order to highlight the region where our cluster members RVs correspond most closely with the 
field star RVs. 
(b) Histogram of the RV membership probabilities of the \ttRV~stars that have been 
observed $\geq3$ times.  We choose to define our RV members as those stars with $P_{RV} >$ 0\%.} 
\label{mems}
\end{figure}

In order to calculate a given star's RV membership probability, we first fit separate Gaussian functions for both the cluster and 
field-star populations to a histogram of the RV measurements for all single stars.   
We then define the  membership probability of a given star by the usual equation:
\begin{equation} \label{memeq}
P_{RV}(v) = \frac{F_{cluster}(v)}{F_{field}(v)+F_{cluster}(v)}  
\end{equation}
where $F_{cluster}(v)$ is the cluster-fit value for RV $v$, and 
$F_{field}(v)$ is the field-fit value for the same RV \citep{vas58}.   We use only single
stars in computing the cluster and field fits; we then re-normalized the Gaussians to the full sample size including velocity-variable stars.  
Figure~\ref{mems}(a) shows the 
RV histogram as well as the fits to the field and cluster stars used in Equation~\ref{memeq}.  The 
cluster is best fit by a Gaussian centered on -42.43 \kms~with $\sigma$ = 0.76 \kms. The cluster fit 
indicates a total 485 stars integrated over the Gaussian fit.  The field 
is best fit by a Gaussian centered on -35.51 \kms~with $\sigma$ = 38.01 \kms~and indicates a total of 314 stars.

As mentioned above, Monte Carlo simulations have indicated that we require three observations 
covering a baseline of at least one year in order to be 95\% confident that a star is either constant or 
variable in velocity (i.e., a single star or a binary- or multiple-star system). 
If this criterion is not met, we classify the star's membership as unknown [U]. 
We have calculated RV membership probabilities for the \ttRV~stars that do meet this criterion.
For single stars we use the average RV to perform this calculation.  For binary 
stars with orbit solutions we use the center-of-mass velocity \citep{gel09}; for binary stars without
orbit solutions, we use the average RV.  For single stars and binaries without orbit solutions 
that have both WIYN and DAO RV measurements, we use a weighted average (weighted by the precision of each 
observatory; see Section~\ref{comb}) to calculate this average RV.  We then round the resulting RV membership 
probability to the nearest integer value and quote this as the $P_{RV}$.
Upon review of the histogram of membership probabilities 
(Figure~\ref{mems}(b)), we choose to define our RV members as those with a non-zero $P_{RV}$.
We note that there are only 31 stars with 0\% $<P_{RV}<$ 80\%. 
NGC 188 is out of the plane, and well separated from the field in RV space.  
As such, in the range of RVs that yield $P_{RV}>$ 0\% we would only expect to find a $\sim$6\% contamination by field 
stars.  This contamination will arise from field stars whose RVs are similar to those of the cluster members.  We have 
expanded this portion of Figure~\ref{mems}(a) and plot the region in the right side of the figure.

We can securely compute RV membership probabilities for single stars as well as binaries with orbit solutions.
For all stars, we also use the P03 PM membership probabilities as a further constraint on cluster membership, 
including only those with non-zero $P_{PM}$ as cluster members.  In addition to our RV membership probabilities,   
we classify single stars as either single members [SM] or single non-members [SN], based on the RV membership 
probability.  Binary stars with orbit solutions are classified as either binary members [BM]
or binary non-members [BN], based on the RV membership probability derived using their center-of-mass velocity \citep{gel09}.
For velocity-variable stars without orbit solutions, our RV membership probabilities are less certain; therefore we simply provide the 
classification as a distinction.  
Velocity-variable stars without orbit solutions are classified as binary likely members [BLM] if $P_{RV}(\overline{RV})>$~0\% 
(where $\overline{RV}$ is the star's average RV), binary unknown [BU] if the range in RV measurements 
includes the cluster mean velocity but $P_{RV}(\overline{RV})=$~0\%, or binary likely non-member [BLN] if all RV measurements 
fall outside of the cluster distribution.  In addition to our stars the have been observed $<$ 3 times, we also classify certain 
stars as unknown [U] that do not have sufficient measurements to yield a reliable class (e.g., rapid rotators with few 
observations).   Note that in Table~\ref{RVtab}, we have only included our 
$P_{RV}$ measurements for the SM, SN, BM, and BN stars, as the RV memberships derived for the remaining classes of stars are uncertain. 
Table~\ref{memnums} displays the number of stars within each classification.

We define our sample of cluster members as the SM, BM and BLM stars; in total, we find \tm~likely member stars. 
Including only the SM and BM cluster members, we find a mean cluster RV of \mrv, with a standard deviation about the mean of  0.64 \kms.

\subsubsection{Comparison of RV and PM Memberships}

We find good agreement between our RV membership probabilities and the P03 PM membership probabilities. 
There are 452 stars within our magnitude limits for which P03 calculates a $P_{PM} \geq$ 80\% .  
We have secure membership probabilities for 384 of these stars (meaning either $\geq$3 observations on a single star 
or an orbit solution for a binary star).  Of these, we find 356 (92\%) to have a $P_{RV} \geq$ 80\%
and 365 (95\%) to have  $P_{RV} >$ 0\%. Of the 19 PM member stars for which we find zero RV membership
probability, 5 lie far from a single star isochrone.  

In total, we find 358 single stars with $P_{RV} \geq$ 80\%.  Of these stars, 
289 (81\%) have $P_{PM} \geq$ 80\% and 340 (94\%) have $P_{PM} >$ 0\%. 
We also find 80 binary stars with orbit solutions to have $P_{RV} \geq$ 80\% \citep{gel09}.
Of these, 70 (88\%) have a $P_{PM} \geq$ 80\%, and 77 (96\%) have a non-zero $P_{PM}$.
There are 25 single stars and binary stars with orbit solutions for which we find $P_{RV} >$ 0\% that P03 find zero 
PM membership probability.  Of these stars, 4 lie far from a single star isochrone.   
As stated above, we expect a $\sim$6\% contamination of field stars
within our $>$0\% RV membership cut.  Indeed, out of our 453 single stars and binary stars with orbit solutions for 
which we find $P_{RV} >$ 0\%, we would expect to find $\sim$25 field stars.

In order to reduce this contamination, we use both the RV and PM membership information to identify cluster members.
Specifically, we have chosen to define our 3D members as those stars that have a non-zero P03 PM membership probability
and non-zero RV membership probability.  These \tm~stars are used in the subsequent analyses.

\subsection{Velocity-Variable Stars} \label{var}

RV variable stars are evident from the standard deviations of their measurements. 
As described in Section~\ref{prec} and shown in Figure~\ref{3all.eoveri}, we have chosen a 
cutoff value of 4 times our precision (\eoi~$>$ 4) to distinguish single stars from velocity-variable stars.  
For stars with both WIYN and DAO measurements, we take a weighted average of the two \eoi~measurements 
resulting from the different precision values\footnote{
$(e/i)^2 = \frac{(N_W-1)*(e/i)_W^2+(N_D-1)*(e/i)_D^2}{N_W+N_D-2}$ ; where $N$ is the number of measurements. The subscript
$W$ corresponds to WIYN while $D$ corresponds to DAO.  Each respective (\eoi)$_{W or D}$ measurement is calculated using the 
appropriate precision value.}.  
If a star should have multiple observations from one 
observatory and only one observation from the other, we exclude that single observation and use 
the other observations with the appropriate precision value in calculating the \eoi~measurement.
Note, we have not included \eoi~measurements in Table~\ref{RVtab} for RR stars or SB2 binaries, as the 
uncertainties on these measurements are not well defined.
In total, we find \tvm~velocity-variable cluster members.  
We assume that our velocity variables are binary- or multiple-star systems, and seek to derive orbit solutions for all of our 
detected binaries.  To date, we have determined orbit solutions for \orb~binaries, 
\BM~of which are probable members; 16 are double-lined and 62 are single-lined \citep{gel09}.  See Table~\ref{memnums} for the 
number of stars within all of our velocity-variable classes.

\subsection{Previously Studied Photometric Variable Stars and X-ray Sources}
We have also examined known photometric variable stars and X-ray sources for membership
and RV variability.  
Table~\ref{phvar} presents the previously studied 
photometric-variable stars (with prefix ``V'' from \citet{zha02} and ``WV'' from \citet{kaf03}) 
that are within our magnitude limits and have $\geq$3 observations.  
Table~\ref{txray} displays similar information for the X-ray sources 
(with prefix ``X'' from \citet{bel98} and ``GX'' from \citet{gon05}).  For both tables, the column headings 
describe the same measurements as in Table~\ref{RVtab}, with the exception of the first column that provides the 
respective photometric variable or X-ray ID.
There are a few photometric variable stars that we have observed, but have not included in Table~\ref{phvar}; each are W UMa stars \citep{zha02}.
We have observed star 5337 (V4) 5 times at the WIYN 3.5m and find an average RV of -45.7 \kms~with a standard deviation of 1.19 
\kms.  We do not observe this star to be a rapid rotator; however the correlation function is unusually shaped requiring more detailed analysis.
We require more observations of this star to derive a reliable membership and classification.
We have also observed stars 5361 (V3), 4989 (V5), and 5209 (V13) multiple times,
but have been unable to derive reliable RV measurements from their spectra, most likely due to their rapid rotation.  

\begin{deluxetable}{l c }
\tabletypesize{\small}
\tablecaption{Number of Stars Within Each Classification\label{memnums}}
\tablehead{\colhead{Class} & \colhead{N stars}}
\startdata
SM & \SM \\
SN & \SN \\
BM & \BM \\
BN & \BN \\
BLM & \BLM \\
BU & \BU \\
BLN & \BLN \\
U & \U \\
\enddata
\end{deluxetable}

\begin{deluxetable*}{l c c c c l l}
\tabletypesize{\small}
\tablewidth{0pt}
\tablecaption{Previously Studied Photometric Variable Stars \label{phvar}}
\tablehead{\colhead{Var} & \colhead{ID$_W$} & \colhead{P$_{RV}$ (\%)} & \colhead{P$_{PM}$ (\%)} & \colhead{\eoi} & \colhead{Class} & \colhead{Comment}}
\startdata
V8, V372 Cep & 5629 & 0 & 0 & - & BN & SB2, RR Secondary, X-ray source (X26,GX1), RS CVn\\
V9 & 4736 & 98 & 97 & 0.59 & SM & G\\
V11 & 4705 & 98 & 98 & - & BM & SB2, X-ray source (GX18), G\\
V12 & 5762 & 66 & 97 & - & BM & SB2, EA\\
WV3 & 5379 & 98 & 98 & 14.96 & BM & SB1, BS\\
WV4 & 4304 & 98 & 97 & 2.14 & SM & \\
WV8 & 4750 & 97 & 96 & 2.91 & SM & \\
WV12 & 4997 & 98 & 85$^a$ & 0.66 & SM & \\
WV19 & 6208 & 0 & 0 & 0.35 & SN & \\
WV24 & 4656 & 97 & 98 & 0.77 & SM & EA\\
WV26 & 3953 & - & 97 & 4.79 & BLM & SB1\\
WV28 & 4508 & - & 98 & 30.30 & BU & SB1, X-ray source (GX20)\\ 
WV33 & 5769 & 0 & 0 & 1.90 & SN & \\
WV34 & 5767 & 98 & 98 & 0.92 & SM & \\
\enddata
\footnotesize
\tablenotetext{a}{
This 85\% PM membership probability is taken from the \citet{din96} study.
The star has a nearby image (within $\sim$5$\arcsec$) in the P03 study,
and therefore Platais et al.~cannot reliably quote a PM membership.  Platais et al.~recommend using 
the \citet{din96} PM membership.}
\end{deluxetable*}

\begin{deluxetable*}{l c c c c l l}
\tabletypesize{\small}
\tablewidth{0pt}
\tablecaption{Previously Studied X-ray Stars \label{txray}}
\tablehead{\colhead{X-ray Source} & \colhead{ID$_W$} & \colhead{P$_{RV}$ (\%)} & \colhead{P$_{PM}$ (\%)} & \colhead{\eoi} & \colhead{Class} & \colhead{Comment}}
\startdata
X1 & 1278 & - & 0 & - & BLN & RR\\
X5 & 1141 & 98 & 88 & 0.59 & SM & G\\
X11 & 483 & - & 0 & - & BU & SB2? \\
X21 & 5258 & - & 92 & - & U & RR \\
X26, GX1 & 5629 & 0 & 0 & - & BN & SB2, RR secondary, RS CVn, photometric variable (V8)\\
X29 & 5027 & 98 & 96 & 1.51 & SM & G, FK Comae-like star \citep[I-1;][]{har85b}\\
GX18 & 4705 & 98 & 98 & - & BM & SB2, G, photometric variable (V11)\\
GX20 & 4508 & - & 98 & 30.30 & BU & SB1, photometric variable (WV28)\\
GX27 & 4230 & - & 93 & - & BU & SB1, RR \\
GX28 & 4289 & 98 & 98 & 71.16 & BM & SB1, G \\
\enddata
\end{deluxetable*}

\section{Discussion}

\subsection{Color-Magnitude Diagram} \label{cmd}

\begin{figure}[!t]
\plotone{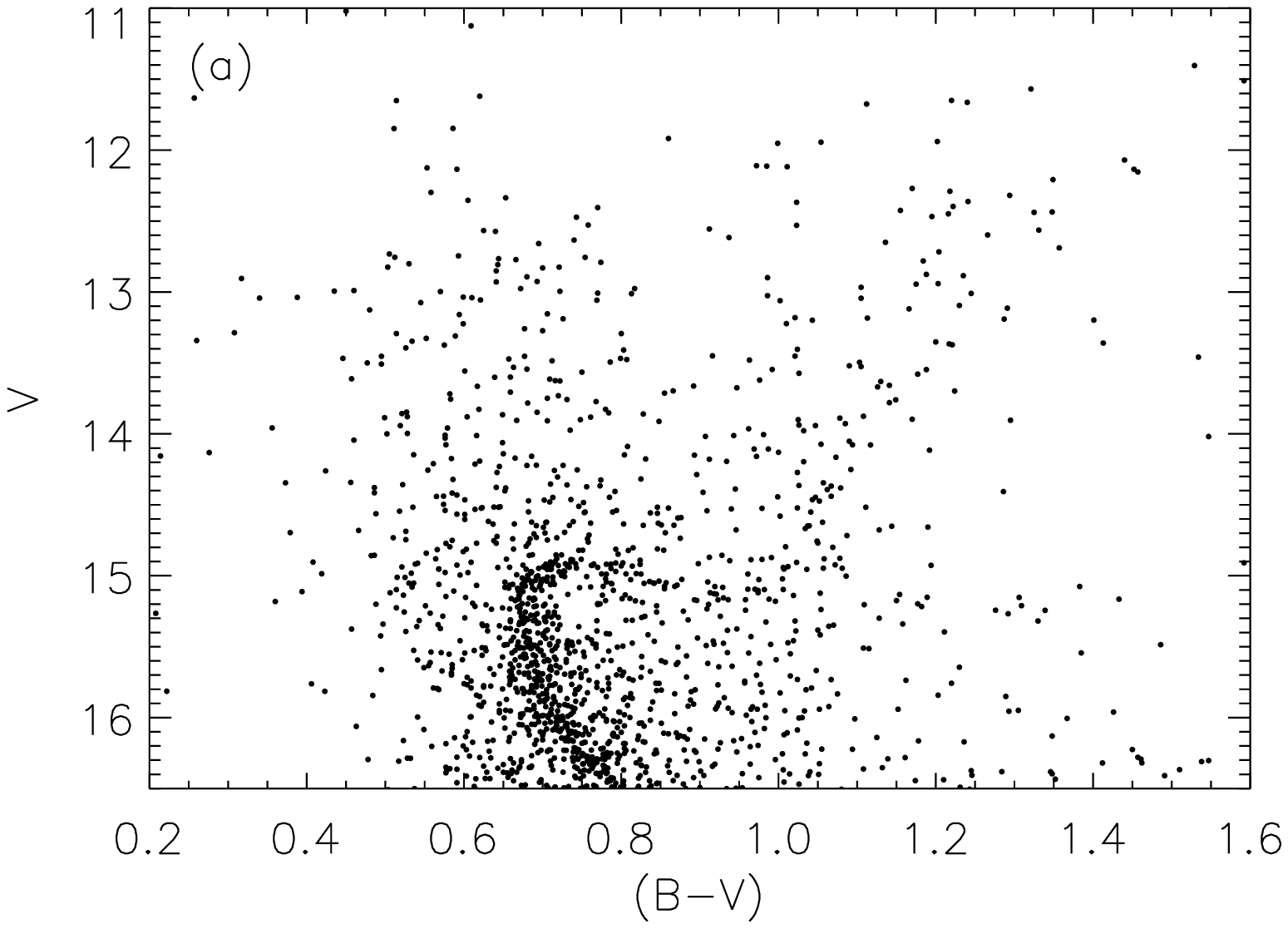}
\plotone{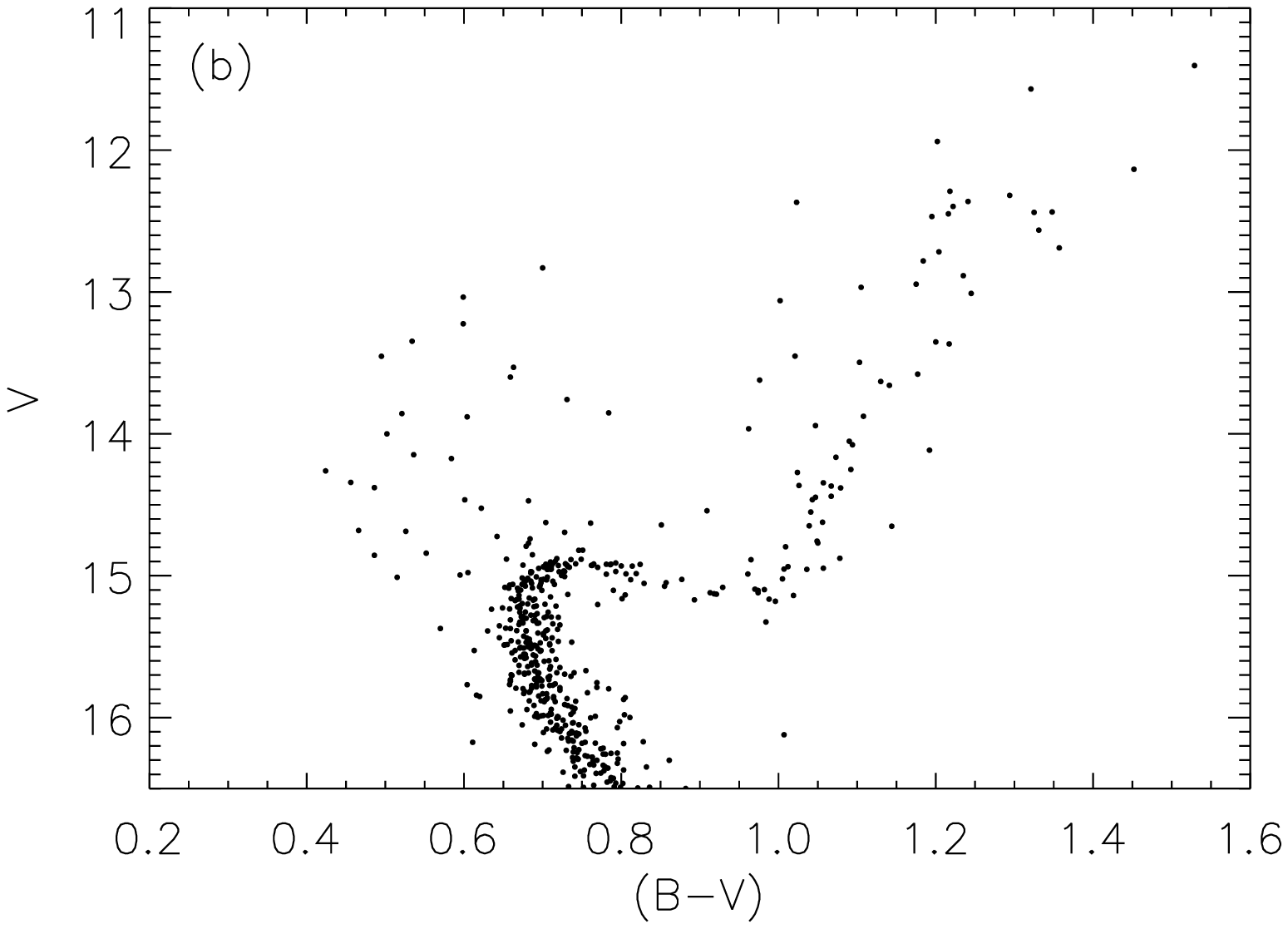}
\plotone{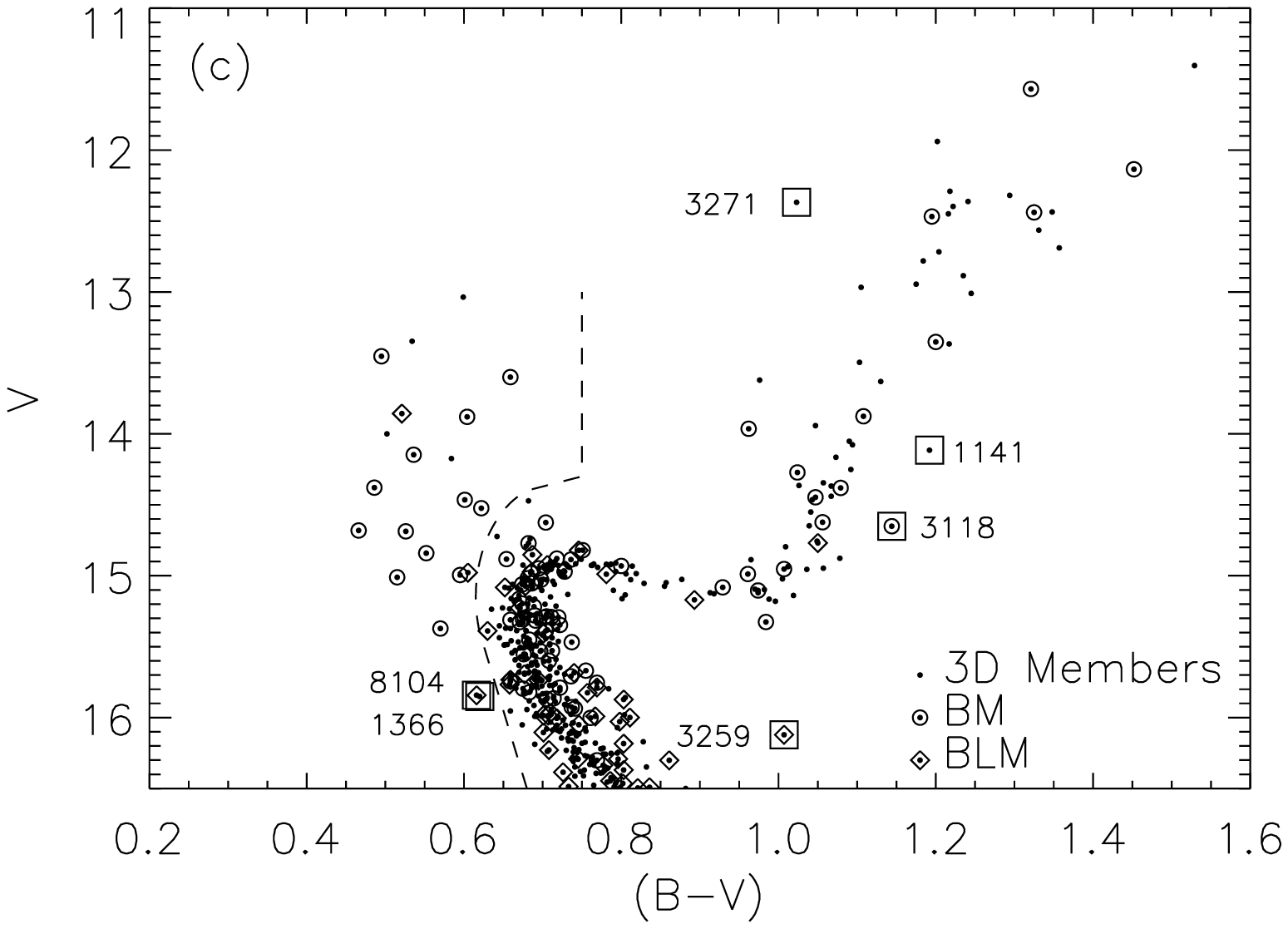}
\caption{\footnotesize (a) CMD of all  stars in the direction of NGC 188 within our sample.
(b) CMD of all such stars whose RV membership is $>$0\%. 
(c) CMD of all such stars whose RV membership is $>$0\% and PM membership is $>$0\%.
There are \tm~3D kinematic members, \BM~binary members [BM], and \BLM~binary likely members [BLM].
BS stars are defined as being bluer than the dashed line.
We have identified 21 BS stars, 16 of which are velocity variables, and 66 red giants, 16 of which
are velocity variables. We have also boxed stars of note.}
\label{CMDs} 
\end{figure}

Without any membership selection, the color-magnitude diagram (CMD) of NGC 188, within our magnitude range and
spatial coverage, reveals little more than a poorly defined main sequence.
Plot (a) of Figure~\ref{CMDs} shows the locations on a CMD of each of the 1498 
stars within our sample of NGC 188.  Plot (b) shows only our RV member stars. 
The power of our combined RV and PM kinematic membership selection is 
clearly shown in plot (c), which displays our cleaned CMD plotting only our \tm~3D kinematic member stars 
($P_{PM} >$~0\% and $P_{RV} >$~0\%), and highlighting our \tvm~velocity-variable member stars.  
The main sequence is clearly visible in plot Figure~\ref{CMDs}(c) along with the 
main-sequence turnoff at about $V \sim$ 14.9 and a well-defined giant branch.  

We have defined our BS stars as being bluer than the dashed line drawn in Figure~\ref{CMDs}(c) and having 
a non-zero PM and RV membership probability.
These stars are both bluer and brighter than the main sequence and main-sequence turnoff. 
We have excluded stars near to the turnoff so as to be sure not to include equal-brightness main-sequence binaries.   
The blue straggler stars are noted as BS in the comment field of Table~\ref{RVtab}.   
We have identified 21 BS stars, 16 of which are velocity variables with \eoi $>$ 4.
Two of our BS stars (1366 and 8104) are fainter than the traditional BS region and are noted below. 
All but two of our BS stars were identified by P03; star 
1947 has $P_{PM}$ = 2\%, and star 1366 has $P_{PM}$ = 3\%.

Interestingly, we note the very large frequency of detected binaries amongst the BS stars (16 out of 21). We will discuss these binaries in detail in our next paper, \citet{gel09}.  Here we simply note that this large frequency is secure as we have derived reliable orbit solutions for all but three of our BS velocity variables; two of these remaining BLM stars have preliminary orbit solutions and are certainly binary stars.

We find 66 red giant stars \footnote{\footnotesize Here, red giants are defined as having 
$(\bv)\geq$~0.95.}, 16 of which are velocity variables.
We also note the large scatter amongst the giant branch stars in our cleaned CMD.  Indeed, the NGC 188 giant branch 
is much broader and less well defined than that of M67, a similarly aged old open cluster (4.5 Gyr) 
\citep[e.g.,][]{jan84}.  
This scatter in the NGC 188 giant branch was previously noted in the photometric studies by \citet{mcc74}, \citet{mcc77}, and \citet{twa78}, 
and is confirmed after our careful membership selection.  The origin is still uncertain.

\subsubsection{Stars of Note}

In this section we provide details on the stars of note that we have boxed in Figure~\ref{CMDs}(c).  
The reader is reminded that we expect $\sim$6\% contamination from field stars within our sample of 
stars with $P_{RV} >$ 0\%.  We have attempted to reduce this contamination by excluding stars with 
$P_{PM}$ = 0\%.  However, there may still be field stars within our cluster member sample. 
We note that all of these stars are located in the outer regions of the cluster.  This may be an 
important clue to their formation mechanisms as well as their subsequent dynamical evolution.  
However, it may also increase the likelihood that they are field stars. 
For most of these stars, the $P_{PM}$ and $P_{RV}$ provide strong evidence for cluster membership.

\paragraph{3271: }
Star 3271 is much bluer than its companions on the giant branch.  Indeed, this star is too blue to reside on the 
horizontal branch or the lower AGB (as compared to a 7 Gyr solar-metallicity Padova isochrone from \citet{gir00}).  This star 
is located at a distance of 15.1 arcmin (6.4 core radii) from the cluster center, and has a $P_{PM}$ = 94\% and a 
$P_{RV}$ = 98\%. We do not detect it as a velocity variable, with an \eoi=0.54, and 17 RV 
measurements.  However, this star may indeed be a long-period or low-amplitude binary that lies outside of our detection 
limits. No combination of two stars from the main sequence through the giant branch can account for this star's position on 
the CMD. We suggest that this star may be a wide red giant - BS binary system. 

\paragraph{1141: }
Star 1141 has a $P_{PM}$ = 88\%, a $P_{RV}$ = 98\%, an \eoi~measurement of 0.59, and is located at a radius of 
17.7 arcmin (7.5 core radii).  This star is redder than its 
counterparts on the giant branch, and we are unable to reproduce its color by the addition of an undetected 
binary companion.  This star is particularly intriguing, as it is also detected as an X-ray source (X5) by \citet{bel98}.  

\paragraph{3118: }
Star 3118 is also redder than its counterparts on the giant branch, and has a high $P_{RV}$ at 88\%, but only a 
marginal $P_{PM}$ at 34\%.  The system is located at a radius of 18.76 arcmin ($\sim$8 core radii) from the cluster 
center.  We detect this star as a SB2 binary and have derived a preliminary orbit with a period of 
12 days and a near circular eccentricity \citep{gel09}.

\paragraph{1366 and 8104: }
Stars 1366 and 8104 are bluer than the main sequence yet fainter than the more ``traditional'' BS population.  
Star 1366 has a $P_{PM}$ = 23\%, a $P_{RV}$ = 96\% and an \eoi~measurement of 1.67 with 3 observations.
This star is located at a radius of 29.6 arcmin (12.6 core radii) from the cluster center.
Star 8104 has a $P_{PM}$ = 91\% and an \eoi~measurement of 8.90 with 7 observations.  This system is located 
at a radius of 14.1 arcmin (6.0 core radii) from the cluster center.  We require more 
observations to derive a reliable RV membership probability, but we have classified this star as a BLM.
BS formation theories that include main-sequence mergers predict BS stars to populate this region of the 
CMD in addition to the more traditional brighter region \citep[e.g.,][]{bai95,str93}.  These 
stars may in fact be main-sequence merger products that now appear bluer than their main sequence counterparts.

\paragraph{3259: }
Star 3259 is found redward of the main sequence and below the subgiant branch.  This star has 
a $P_{PM}$ = 71\% and an \eoi~measurement of 17.52 with 6 observations.  
We require more observations to derive a reliable RV membership probability, but we have classified this star as a BLM.
3259 is located at a radius of 15.6 arcmin (6.6 core radii) from the cluster center.
Stars found in this region of the CMD are hard to explain through traditional theories of stellar evolution.
No simple photometric combination of cluster members can account for this star's 
position on the CMD.  Though not displayed on Figure~\ref{CMDs}(c),  star 4989 ($V$=16.10, $\bv$=0.97) is a 95\% PM member 
and is located at a radius of 1.8 arcmin (0.8 core radii) from the cluster center.  
This star is identified as a W UMa (V5) by \citet{zha02}.
We are unable to derive reliable RVs for this star from our seven observations (the CCF peak heights fall below our 0.4 cutoff).
However, a visual inspection of the spectra suggests that star 4989 is likely a  RR.
These stars may be analogous to the sub-subgiants discovered in M67 by \citet{mat03}.

\subsection{Spatial Distribution and Mass Segregation} \label{spat}

\begin{figure}[!t]
\plotone{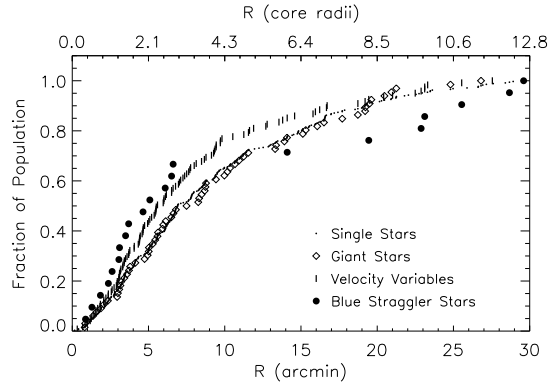}
\caption{\footnotesize Cumulative spatial distribution of our 3D member stars.  
The single, BS, giant, and velocity-variable star populations are compared.  
The velocity-variable stars are distinct from the single stars with 92\% confidence, providing 
evidence for the presence of mass segregation in NGC 188.}
\label{kscum} 
\end{figure}

The currently accepted model predicts that as two-body interactions lead towards equipartition of kinetic energy, we should, 
in a relaxed cluster, expect the more massive stars and star systems to have fallen to the center of the cluster potential 
dominated by the less massive stars.  Mass segregation has been observed in relaxed clusters of a variety of ages, 
from the young open clusters Pleiades \citep[100 Myr;][]{rab98a} and Preaesepe \citep[800 Myr;][]{rab98b} to the old open cluster M67 
\citep[4.5 Gyr;][]{mat86}.  \citet{har85a} derived a relaxation time for NGC 188 of $\sim$10$^8$
years, which suggests that NGC 188 may have lived through at least tens of relaxation times.  Thus we would expect 
NGC 188 to be mass segregated.   Indeed \citet{sar99} found evidence for mass segregation in their photometric study of NGC 188.   
In addition, a study by \citet{kaf03} found their NGC 188 photometric variable star sample to be centrally concentrated, 
affirming the presence of mass segregation if their photometric variables are binary stars.  
Conversely \citet{din96} did not find their BS population to show significantly more central concentration than the giant stars.
If we assume that BS stars are formed through mergers of two (or more) stars \citep[e.g.,][]{bai95,str93}, we would expect the BS stars
to be more massive than the giants.  Hence this finding suggests a lack of mass segregation amongst the BS 
population, which would appear to disagree with \citet{sar99} and \citet{kaf03}.  
Therefore, we reanalyze NGC 188 for the presence of mass segregation in light of the precise membership 
determination presented in this paper.  We compare the spatial distributions of the giant, velocity variable, and BS stars 
to that of the single stars (i.e., stars with constant velocity) within our NGC 188 sample, and plot the results in Figure~\ref{kscum}. 
 
We first investigate the velocity variables by comparing their spatial distribution to the single stars.
A Kolmogorov-Smirnov (K-S) test shows that the velocity variables are derived from a different parent spatial distribution 
than the single stars with 92\% confidence.  Presuming that the velocity variables are binary- or multiple-star systems,
then the central concentration of the more massive binary star population offers further evidence for the presence of mass 
segregation in NGC 188.  

Next, we investigate the spatial distribution of the NGC 188 BS population by again comparing to the single star population.
A K-S test shows a low distinction of 63\% between the two populations.  We therefore do not find the full population of 
BS stars to be centrally concentrated with respect to the single stars, agreeing with \citet{din96}.
The majority of our BS stars are found in binary systems; additionally, as stated above, the formation of BS stars is 
likely the result of mergers of two (or more) stars.  Hence their combined mass is likely much greater than that of a 
single star in our sample.  As such, this lack of central concentration is surprising and may provide a significant insight 
into the formation mechanisms and subsequent dynamical evolution of the NGC 188 BS stars.  

Interestingly, in the similarly aged open cluster M67 (4.5 Gyr), \citet{mat86} find secure evidence for mass segregation of 
the BS population.  Perhaps importantly, their sample only extended to $\sim$5 core radii\footnote{assuming a core radius 
of 1.3 pc as derived by \citet{zha96}}, whereas our NGC 188 sample extends to $\sim$13 core radii.  
If we limit our NGC 188 sample to 5 core radii and again compare the spatial 
distribution of the BS and single star populations, we find 95\% distinction between the two 
populations.  Thus, this restricted population of BS stars does show evidence for central concentration.

We also note the seemingly bimodal spatial distribution amongst the NGC 188 BS stars.
We suggest that we may be observing two populations of BS stars, with a centrally 
concentrated population, within $\sim$5 core radii, and a halo population extending to the edge of the cluster.  
Similar bimodal BS spatial distributions have been observed in some globular clusters (e.g., M55 \citep{lan07}, 47 Tuc \citep{fer04}, 
NGC 6752 \citep{sab04}, and M3 \citep{fer97}).
These bimodal spatial distribution are likely the result of the formation mechanism(s) and dynamical evolution 
of the BS populations. 
Scattering experiments performed by \citet{leo92} suggest that BS stars formed through collisions within a population of 
primarily short-period binaries may be given enough momentum to achieve long radial orbits, spending the majority of their 
time in the outer cluster regions where the relaxation time is longer.  Thus these halo BS stars may not attain significant 
central concentration. This result agrees with \citet{sig94} who suggest that the BS stars are likely formed in the core and 
ejected to varying cluster radii depending on the recoil of the dynamical interaction.  BS stars that are only ejected to a few 
core radii quickly fall back to the cluster center through dynamical friction and mass segregation processes, while those ejected 
farther may remain as a halo population.  Conversely Ferraro et al. (1997, 2004) suggest that such a bimodal distribution is 
the result of two BS formation mechanisms, with the outer BS stars formed through mass transfer in primordial 
binaries, and the inner BS stars formed from stellar interactions which lead to mergers.

\subsection{Cluster Radial-Velocity Dispersion} \label{disp}

We can use Equation~\ref{scm2} and our detailed analysis of the precision of our measurements to solve for
the combined RV dispersion of NGC 188  : 
$\sigma_{cb}^2 = \sigma_c^2 + \sigma_b^2 = \sigma_{obs}^2 - \sigma_i^2$.  
Using the average values derived above, the resulting global value of the combined velocity dispersion is 
$\sigma_{cb}$ = \vderr.  This value is likely inflated due to the 
presence of undetected binaries; we attempt to evaluate this overestimate below.

Mathieu (1983, 1985) studied the effect of such undetected binaries in the old open cluster M67 
through a Monte Carlo analysis.  
This cluster has a 
similar observed binary frequency and combined velocity dispersion to NGC 188.  The Monte Carlo simulations used an \citet{abt76} 
binary distribution.  The results from that study suggest that for a combined velocity dispersion of 0.5~\kms~and
a total binary fraction of 50\%, one would expect the combined velocity dispersion to be inflated by 0.23~\kms.  Given our 
observed hard binary (P$<$ 10$^4$) fraction of roughly 30\%, we expect a contribution from undetected binaries to
be similar to the values derived by Mathieu (1983, 1985).

Given the mean velocity dispersion, we can estimate the virial mass of NGC 188.  We use the equation given by \citet{spi87}, 
in projection:
\begin{equation} \label{mvirial}
M = \frac{10\langle \sigma_c^2\rangle R_h}{G}
\end{equation}
where $\sigma_c$ is the line-of-sight (radial) velocity dispersion, and $R_h$ is the projected half-mass radius.  We use the P03 PM 
members (i.e., stars with $P_{PM} >$ 0\%) to derive $R_h$ = 5.8 $\pm$ 0.3 pc.  
We estimate $\sigma_c$ = 0.41 $\pm$ 0.04 \kms, using the correction value derived 
by  Mathieu (1983, 1985) to recover $\sigma_c$ from our derived $\sigma_{cb}$ given above, and assuming an error equal to that of
$\sigma_{cb}$.
The resulting virial mass is 2300 $\pm$ 460 \Msolar.  This value agrees well with the total mass one gets by counting 
the P03 PM members; if we assume a mass-to-light relationship of $M \propto L^{1/3}$ and use the proper bolometric correction factors, 
we find a total mass of 2350 \Msolar.  
Our mass estimate also lies within the 90\% confidence range of 230 to 2600 \Msolar derived by \citet{har85a}
with an isotropic King model fit \citep{kin66}.  Additionally, our mass estimate agrees with the total mass of 3800 $\pm$ 1600
\Msolar derived by \citet{bon05} from integrating an extrapolated mass function.  

\begin{figure}[!t]
\plotone{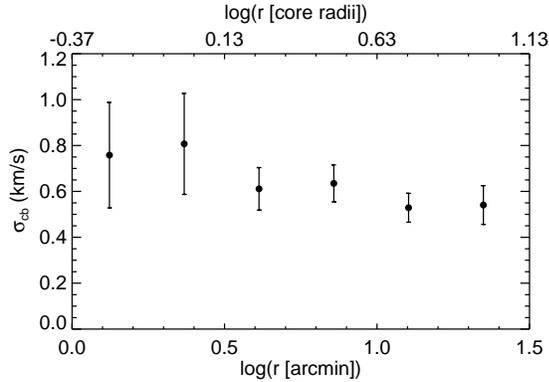}
\caption{\footnotesize Combined RV dispersion of single cluster members in NGC 188 plotted as a function of radius from the 
cluster center.  We have plotted only our SM stars, and use only the data obtained at the WIYN 3.5m.
The radial distances plotted represent the average values in the log for each bin. The measurement errors are from sampling.  The central combined RV dispersion values are likely 
inflated by undetected binaries.  Correcting the central bins for undetected binaries yields a distribution that is 
indistinguishable from an isothermal distribution.}
\label{vdisp} 
\end{figure}

We have also looked for a variation in the combined RV dispersion with radial distance from the cluster center,
again using only our SM stars ($P_{RV} >$ 0\%, $P_{PM} >$ 0\% and \eoi~$<$ 4).
We have chosen to plot the upper limits on the combined RV dispersion without attempting to correct for undetected 
binaries, and have binned the data in equal steps in log-radius.  We only use the WIYN RV data for this analysis to 
avoid unnecessary complications from the difference in precision of the two data sets.  The results are shown in 
Figure~\ref{vdisp}. The combined RV dispersion ranges from $\sim$0.8 \kms~in the radial bins closest to the cluster center to 
$\sim$0.6 \kms~in the bins farthest away from the cluster center.  We note that the central rise in combined RV dispersion 
is likely inflated due to undetected binaries, as the binaries are centrally concentrated in NGC 188 (see Section~\ref{spat}).  
Indeed, applying the correction described above to the central bins would lower
these values to approximately that of our farthest bins, where we would expect the least inflation by undetected 
binaries.  Moreover, this correction results in an RV dispersion distribution that is indistinguishable from an 
isothermal distribution.      

\citet{har85a} also performed a study of the radial dependence of the combined RV dispersion of 
NGC 188, though with a smaller sample size than our study.  They found a higher combined dispersion in the inner 
cluster region  of 1.16 (with a range of 0.88-1.81) \kms, but the values in their remaining radial bins, of 0.51 (0.22-1.06) \kms~at 
5.6-13 arcmin and 0.20 (0.12-0.40) \kms~at 13-20 arcmin, are similar to the values we have obtained here.
Undoubtedly, the combined dispersion measured by Harris \& McClure
in the inner bin was inflated due to undetected binaries,
as that paper also concluded based on Monte Carlo models of
observations of binaries.

\section{Summary}

We have presented a detailed discussion of our RV study of the rich old open cluster 
NGC 188.  As the first paper in a series of WIYN Open Cluster Study papers presenting precise stellar radial-velocity surveys, we have provided an in-depth 
description of our data reduction process and paid particular care in establishing our measurement precision of 0.4 \kms.  For the first time, full 3D kinematic membership 
probabilities have been used to study the cluster.  We have thus been highly successful in eliminating 
field star contamination within our sample, finding \tm~probable member stars. Importantly, \tvm~of 
these are velocity-variable stars, providing an extensive population of dynamically hard binaries to be studied.

We have used our secure member stars to produce a cleaned CMD that displays our population of BS stars, 
76\% of which are binaries.  We also note the large scatter within 
the giant branch.  In addition, we have discussed six stars of note that lie far from a standard single-star isochrone. 
Stars 3271, 1141 and 3118 are located near 
the giant branch.  Stars 1366 and 8104 are BS stars that lie below the traditional brighter blue straggler region. 
Star 3259 may be analogous to the sub-subgiants of M67 \citep{mat03}.

We have also investigated the spatial distribution of the stellar populations within the cluster.    
We find the binary stars in our sample to be central concentrated with 92\% confidence, providing further
evidence for the presence of mass segregation within NGC 188.  The BS spatial distribution appears to be bimodal, and 
we find no statistical evidence that the full BS population is centrally concentrated.  This finding is surprising and may 
suggest that we are observing two populations of BS stars in NGC 188, including a centrally concentrated distribution as well 
as a halo BS population.
Further study is necessary to determine whether this bimodal distribution is a sign of two distinct 
formation mechanisms, or a result of dynamical scattering during the BS formation.

Our comprehensive error analysis has allowed us to study the RV dispersion of the cluster, and we find an upper 
limit on our mean combined RV dispersion of \vderr.  The results from Mathieu (1983, 1985)
suggest that this velocity dispersion may be inflated by as much as 0.23 \kms~from undetected binaries.  
We have used this corrected velocity dispersion to estimate a mass of 2300 $\pm$ 460 \Msolar in NGC 188, which agrees 
well with the value derived by counting the mass in P03 PM members, as well as other NGC 188 mass estimates from the literature.
The combined RV dispersion decreases with increasing cluster radius; however the dispersion values in the central bins 
are likely inflated due to undetected binaries.  Correcting the central bins for undetected binaries yields 
a RV dispersion distribution that is indistinguishable from an isothermal distribution.

The WIYN Open Cluster Study will continue its survey of NGC 188 in order to provide membership 
data on all stars in our complete sample.  We are currently in the process of analyzing our binary population
for orbital solutions, and seek to identify all RV variables in the cluster within our observational constraints.
With these orbit solutions, we will study the binary distribution in period, eccentricity, and mass ratio, and our 
identified RV variables will allow us to constrain the cluster binary fraction.
As one of the oldest and most widely studied open clusters, NGC 188 holds the key to understanding 
the long term evolution of binary systems, merger products, blue stragglers, and stellar dynamics in clusters with 
rich binary populations.

\acknowledgments
The authors would like to express their gratitude to the staff of the WIYN Observatory without whom we would 
not have been able to acquire these thousands of superb stellar spectra.  Many thanks to Sydney Barnes and S\o ren 
Meibom who both played a significant part in the development of our error budget analysis in Section~\ref{erbud}. 
We also want to thank the many undergraduate and graduate students who have obtained spectra over the years at WIYN
for this project.
We would also like to acknowledge R. F. 
Griffin and J. E. Gunn for contributing their NGC 188 RVs to our project.  In turn, they wish to express their 
thanks to the Palomar Observatory for the use of the 5m telescope.  Thanks to Murray Fletcher for his expertise 
in developing the DAO RVS instrument.  We also thank Jim Hesser who acquired 
a portion of the DAO NGC 188 data.  Finally, we would like to express our gratitude to the referee Ted von Hippel for 
his many insightful comments.  This work was funded by the National Science Foundation grant AST-0406615 and 
the Wisconsin Space Grant Consortium.

Facilities: \facility{WIYN 3.5m}, \facility{DAO 1.2m}, \facility{Palomar 5m}

\end{document}